 \input pictex.tex   
\immediate\write10{Package DCpic 2002/05/16 v4.0}

\catcode`!=11 

\newcount\aux%
\newcount\auxa%
\newcount\auxb%
\newcount\m%
\newcount\n%
\newcount\x%
\newcount\y%
\newcount\xl%
\newcount\yl%
\newcount\d%
\newcount\dnm%
\newcount\xa%
\newcount\xb%
\newcount\xmed%
\newcount\xc%
\newcount\xd%
\newcount\ya%
\newcount\yb%
\newcount\ymed%
\newcount\yc%
\newcount\yd
\newcount\expansao%
\newcount\tipografo
\newcount\distanciaobjmor
\newcount\tipoarco
\newif\ifpara%
\newbox\caixa%
\newbox\caixaaux%
\newif\ifnvazia%
\newif\ifvazia%
\newif\ifcompara%
\newif\ifdiferentes%
\newcount\xaux%
\newcount\yaux%
\newcount\guardaauxa%
\newcount\alt%
\newcount\larg%
\newcount\prof%
\newcount\auxqx
\newcount\auxqy
\newif\ifajusta%
\newif\ifajustadist
\def\objPartida{}%
\def\objChegada{}%
\def\objNulo{}%


\def\!vazia{:}

\def\!pilhanvazia#1{\let\arg=#1%
\if:\arg\ \nvaziafalse\vaziatrue \else \nvaziatrue\vaziafalse\fi}

\def\!coloca#1#2{\edef\pilha{#1.#2}}

\def\!guarda(#1)(#2,#3)(#4,#5,#6){\def\id{#1}%
\xaux=#2%
\yaux=#3%
\alt=#4%
\larg=#5%
\prof=#6%
}

\def\!topaux#1.#2:{\!guarda#1}
\def\!topo#1{\expandafter\!topaux#1}

\def\!popaux#1.#2:{\def\pilha{#2:}}
\def\!retira#1{\expandafter\!popaux#1}

\def\!comparaaux#1#2{\let\argA=#1\let\argB=#2%
\ifx\argA\argB\comparatrue\diferentesfalse\else\comparafalse\diferentestrue\fi}

\def\!compara#1#2{\!comparaaux{#1}{#2}}

\def\!absoluto#1#2{\n=#1%
  \ifnum \n > 0
    #2=\n
  \else
    \multiply \n by -1
    #2=\n
  \fi}


\def\solidline{2}
\def\injectionarrow{3}


\def\atleft{1}

\def\commdiag{0}


\def\!ajusta#1#2#3#4#5#6{\aux=#5%
  \let\auxobj=#6%
  \ifcase \tipografo    
    \ifnum\number\aux=10 
      \ajustadisttrue 
    \else
      \ajustadistfalse  
    \fi
  \else  
   \ajustadistfalse
  \fi
  \ifajustadist
   \let\pilhaaux=\pilha%
   \loop%
     \!topo{\pilha}%
     \!retira{\pilha}%
     \!compara{\id}{\auxobj}%
     \ifcompara\nvaziafalse \else\!pilhanvazia\pilha \fi%
     \ifnvazia%
   \repeat%
   \let\pilha=\pilhaaux%
   \ifvazia%
    \ifdiferentes%
     \larg=1310720
     \prof=655360%
     \alt=655360%
    \fi%
   \fi%
   \divide\larg by 131072
   \divide\prof by 65536
   \divide\alt by 65536
   \ifnum\number\y=\number\yl
    \advance\larg by 3
    \ifnum\number\larg>\aux
     #5=\larg
    \fi
   \else
    \ifnum\number\x=\number\xl
     \ifnum\number\yl>\number\y
      \ifnum\number\alt>\aux
       #5=\alt
      \fi
     \else
      \advance\prof by 5
      \ifnum\number\prof>\aux
       #5=\prof
      \fi
     \fi
    \else
     \auxqx=\x
     \advance\auxqx by -\xl
     \!absoluto{\auxqx}{\auxqx}%
     \auxqy=\y
     \advance\auxqy by -\yl
     \!absoluto{\auxqy}{\auxqy}%
     \ifnum\auxqx>\auxqy
      \ifnum\larg<10
       \larg=10
      \fi
      \advance\larg by 3
      #5=\larg
     \else
      \ifnum\yl>\y
       \ifnum\larg<10
        \larg=10
       \fi
      \advance\alt by 6
       #5=\alt
      \else
      \advance\prof by 11
       #5=\prof
      \fi
     \fi
    \fi
   \fi
\fi} 

\def\!raiz#1#2{\n=#1%
  \m=1%
  \loop
    \aux=\m%
    \advance \aux by 1%
    \multiply \aux by \aux%
    \ifnum \aux < \n%
      \advance \m by 1%
      \paratrue%
    \else\ifnum \aux=\n%
      \advance \m by 1%
      \paratrue%
       \else\parafalse%
       \fi
    \fi
  \ifpara%
  \repeat
#2=\m}

\def\!ucoord#1#2#3#4#5#6#7{\aux=#2%
  \advance \aux by -#1%
  \multiply \aux by #4%
  \divide \aux by #5%
  \ifnum #7 = -1 \multiply \aux by -1 \fi%
  \advance \aux by #3%
#6=\aux}

\def\!quadrado#1#2#3{\aux=#1%
  \advance \aux by -#2%
  \multiply \aux by \aux%
#3=\aux}

\def\!distnomemor#1#2#3#4#5#6{\setbox0=\hbox{#5}%
  \aux=#1
  \advance \aux by -#3
  \ifnum \aux=0
     \aux=\wd0 \divide \aux by 131072
     \advance \aux by 3
     #6=\aux
  \else
     \aux=#2
     \advance \aux by -#4
     \ifnum \aux=0
        \aux=\ht0 \advance \aux by \dp0 \divide \aux by 131072
        \advance \aux by 3
        #6=\aux%
     \else
     #6=3
     \fi
   \fi
}

\def\begindc#1{\!ifnextchar[{\!begindc{#1}}{\!begindc{#1}[30]}}
\def\!begindc#1[#2]{\beginpicture 
  \let\pilha=\!vazia
  \setcoordinatesystem units <1pt,1pt>
  \expansao=#2
  \ifcase #1
    \distanciaobjmor=10
    \tipoarco=0         
    \tipografo=0        
  \or
    \distanciaobjmor=2
    \tipoarco=0         
    \tipografo=1        
  \or
    \distanciaobjmor=1
    \tipoarco=2         
    \tipografo=2        
  \or
    \distanciaobjmor=8
    \tipoarco=0         
    \tipografo=3        
  \or
    \distanciaobjmor=8
    \tipoarco=2         
    \tipografo=4        
  \fi}

\def\enddc{\endpicture}

\def\mor{%
  \!ifnextchar({\!morxy}{\!morObjA}}
\def\!morxy(#1,#2){%
  \!ifnextchar({\!morxyl{#1}{#2}}{\!morObjB{#1}{#2}}}
\def\!morxyl#1#2(#3,#4){%
  \!ifnextchar[{\!mora{#1}{#2}{#3}{#4}}{\!mora{#1}{#2}{#3}{#4}[\number\distanciaobjmor,\number\distanciaobjmor]}}%
\def\!morObjA#1{%
 \let\pilhaaux=\pilha%
 \def\objPartida{#1}%
 \loop%
    \!topo\pilha%
    \!retira\pilha%
    \!compara{\id}{\objPartida}%
    \ifcompara \nvaziafalse \else \!pilhanvazia\pilha \fi%
   \ifnvazia%
 \repeat%
 \ifvazia%
  \ifdiferentes%
   Error: Incorrect label specification%
   \xaux=1%
   \yaux=1%
  \fi%
 \fi%
 \let\pilha=\pilhaaux%
 \!ifnextchar({\!morxyl{\number\xaux}{\number\yaux}}{\!morObjB{\number\xaux}{\number\yaux}}}
\def\!morObjB#1#2#3{%
  \x=#1
  \y=#2
 \def\objChegada{#3}%
 \let\pilhaaux=\pilha%
 \loop
    \!topo\pilha %
    \!retira\pilha%
    \!compara{\id}{\objChegada}%
    \ifcompara \nvaziafalse \else \!pilhanvazia\pilha \fi
   \ifnvazia
 \repeat
 \ifvazia
  \ifdiferentes%
   Error: Incorrect label specification
   \xaux=\x%
   \advance\xaux by \x%
   \yaux=\y%
   \advance\yaux by \y%
  \fi
 \fi
 \let\pilha=\pilhaaux
 \!ifnextchar[{\!mora{\number\x}{\number\y}{\number\xaux}{\number\yaux}}{\!mora{\number\x}{\number\y}{\number\xaux}{\number\yaux}[\number\distanciaobjmor,\number\distanciaobjmor]}}
\def\!mora#1#2#3#4[#5,#6]#7{%
  \!ifnextchar[{\!morb{#1}{#2}{#3}{#4}{#5}{#6}{#7}}{\!morb{#1}{#2}{#3}{#4}{#5}{#6}{#7}[1,\number\tipoarco] }}
\def\!morb#1#2#3#4#5#6#7[#8,#9]{\x=#1%
  \y=#2%
  \xl=#3%
  \yl=#4%
  \multiply \x by \expansao%
  \multiply \y by \expansao%
  \multiply \xl by \expansao%
  \multiply \yl by \expansao%
  \!quadrado{\number\x}{\number\xl}{\auxa}%
  \!quadrado{\number\y}{\number\yl}{\auxb}%
  \d=\auxa%
  \advance \d by \auxb%
  \!raiz{\d}{\d}%
  \auxa=#5
  \!compara{\objNulo}{\objPartida}%
  \ifdiferentes
   \!ajusta{\x}{\xl}{\y}{\yl}{\auxa}{\objPartida}%
   \ajustatrue
   \def\objPartida{}
  \fi
  \guardaauxa=\auxa
  \!ucoord{\number\x}{\number\xl}{\number\x}{\auxa}{\number\d}{\xa}{1}%
  \!ucoord{\number\y}{\number\yl}{\number\y}{\auxa}{\number\d}{\ya}{1}%
  \auxa=\d%
  \auxb=#6
  \!compara{\objNulo}{\objChegada}%
  \ifdiferentes
   \!ajusta{\x}{\xl}{\y}{\yl}{\auxb}{\objChegada}%
   \def\objChegada{}
  \fi
  \advance \auxa by -\auxb%
  \!ucoord{\number\x}{\number\xl}{\number\x}{\number\auxa}{\number\d}{\xb}{1}%
  \!ucoord{\number\y}{\number\yl}{\number\y}{\number\auxa}{\number\d}{\yb}{1}%
  \xmed=\xa%
  \advance \xmed by \xb%
  \divide \xmed by 2
  \ymed=\ya%
  \advance \ymed by \yb%
  \divide \ymed by 2
  \!distnomemor{\number\x}{\number\y}{\number\xl}{\number\yl}{#7}{\dnm}%
  \!ucoord{\number\y}{\number\yl}{\number\xmed}{\number\dnm}{\number\d}{\xc}{-#8}%
  \!ucoord{\number\x}{\number\xl}{\number\ymed}{\number\dnm}{\number\d}{\yc}{#8}%
\ifcase #9  
  \arrow <4pt> [.2,1.1] from {\xa} {\ya} to {\xb} {\yb}
\or  
  \setdashes
  \arrow <4pt> [.2,1.1] from {\xa} {\ya} to {\xb} {\yb}
  \setsolid
\or  
  \setlinear
  \plot {\xa} {\ya}  {\xb} {\yb} /
\or  
  \auxa=\guardaauxa
  \advance \auxa by 3%
 \!ucoord{\number\x}{\number\xl}{\number\x}{\number\auxa}{\number\d}{\xa}{1}%
 \!ucoord{\number\y}{\number\yl}{\number\y}{\number\auxa}{\number\d}{\ya}{1}%
 \!ucoord{\number\y}{\number\yl}{\number\xa}{3}{\number\d}{\xd}{-1}%
 \!ucoord{\number\x}{\number\xl}{\number\ya}{3}{\number\d}{\yd}{1}%
  \arrow <4pt> [.2,1.1] from {\xa} {\ya} to {\xb} {\yb}
  \circulararc -180 degrees from {\xa} {\ya} center at {\xd} {\yd}
\or  
  \auxa=3
 \!ucoord{\number\y}{\number\yl}{\number\xa}{\number\auxa}{\number\d}{\xmed}{-1}%
 \!ucoord{\number\x}{\number\xl}{\number\ya}{\number\auxa}{\number\d}{\ymed}{1}%
 \!ucoord{\number\y}{\number\yl}{\number\xa}{\number\auxa}{\number\d}{\xd}{1}%
 \!ucoord{\number\x}{\number\xl}{\number\ya}{\number\auxa}{\number\d}{\yd}{-1}%
  \arrow <4pt> [.2,1.1] from {\xa} {\ya} to {\xb} {\yb}
  \setlinear
  \plot {\xmed} {\ymed}  {\xd} {\yd} /
\fi
\auxa=\xl
\advance \auxa by -\x%
\ifnum \auxa=0 
  \put {#7} at {\xc} {\yc}
\else
  \auxb=\yl
  \advance \auxb by -\y%
  \ifnum \auxb=0 \put {#7} at {\xc} {\yc}
  \else 
    \ifnum \auxa > 0 
      \ifnum \auxb > 0
        \ifnum #8=1
          \put {#7} [rb] at {\xc} {\yc}
        \else 
          \put {#7} [lt] at {\xc} {\yc}
        \fi
      \else
        \ifnum #8=1
          \put {#7} [lb] at {\xc} {\yc}
        \else 
          \put {#7} [rt] at {\xc} {\yc}
        \fi
      \fi
    \else
      \ifnum \auxb > 0 
        \ifnum #8=1
          \put {#7} [rt] at {\xc} {\yc}
        \else 
          \put {#7} [lb] at {\xc} {\yc}
        \fi
      \else
        \ifnum #8=1
          \put {#7} [lt] at {\xc} {\yc}
        \else 
          \put {#7} [rb] at {\xc} {\yc}
        \fi
      \fi
    \fi
  \fi
\fi
}

\def\modifplot(#1{\!modifqcurve #1}
\def\!modifqcurve(#1,#2){\x=#1%
  \y=#2%
  \multiply \x by \expansao%
  \multiply \y by \expansao%
  \!start (\x,\y)
  \!modifQjoin}
\def\!modifQjoin(#1,#2)(#3,#4){\x=#1%
  \y=#2%
  \xl=#3%
  \yl=#4%
  \multiply \x by \expansao%
  \multiply \y by \expansao%
  \multiply \xl by \expansao%
  \multiply \yl by \expansao%
  \!qjoin (\x,\y) (\xl,\yl)             
  \!ifnextchar){\!fim}{\!modifQjoin}}
\def\!fim){\ignorespaces}

\def\setaxy(#1{\!pontosxy #1}
\def\!pontosxy(#1,#2){%
  \!maispontosxy}
\def\!maispontosxy(#1,#2)(#3,#4){%
  \!ifnextchar){\!fimxy#3,#4}{\!maispontosxy}}
\def\!fimxy#1,#2){\x=#1%
  \y=#2
  \multiply \x by \expansao
  \multiply \y by \expansao
  \xl=\x%
  \yl=\y%
  \aux=1%
  \multiply \aux by \auxa%
  \advance\xl by \aux%
  \aux=1%
  \multiply \aux by \auxb%
  \advance\yl by \aux%
  \arrow <4pt> [.2,1.1] from {\x} {\y} to {\xl} {\yl}}

\def\cmor#1 #2(#3,#4)#5{%
  \!ifnextchar[{\!cmora{#1}{#2}{#3}{#4}{#5}}{\!cmora{#1}{#2}{#3}{#4}{#5}[0] }}
\def\!cmora#1#2#3#4#5[#6]{%
  \ifcase #2
      \auxa=0
      \auxb=1
    \or
      \auxa=0
      \auxb=-1
    \or
      \auxa=1
      \auxb=0
    \or
      \auxa=-1
      \auxb=0
    \fi  
  \ifcase #6  
    \modifplot#1
    \setaxy#1
  \or  
    \setdashes
    \modifplot#1
    \setaxy#1
    \setsolid
  \or  
    \modifplot#1
  \fi  
  \x=#3%
  \y=#4%
  \multiply \x by \expansao%
  \multiply \y by \expansao%
  \put {#5} at {\x} {\y}}

\def\obj(#1,#2){%
  \!ifnextchar[{\!obja{#1}{#2}}{\!obja{#1}{#2}[Nulo]}}
\def\!obja#1#2[#3]#4{%
  \!ifnextchar[{\!objb{#1}{#2}{#3}{#4}}{\!objb{#1}{#2}{#3}{#4}[1]}}
\def\!objb#1#2#3#4[#5]{%
  \x=#1%
  \y=#2%
  \def\!pinta{\normalsize$\bullet$}
  \def\!nulo{Nulo}%
  \def\!arg{#3}%
  \!compara{\!arg}{\!nulo}%
  \ifcompara\def\!arg{#4}\fi%
  \multiply \x by \expansao%
  \multiply \y by \expansao%
  \setbox\caixa=\hbox{#4}%
  \!coloca{(\!arg)(#1,#2)(\number\ht\caixa,\number\wd\caixa,\number\dp\caixa)}{\pilha}%
  \auxa=\wd\caixa \divide \auxa by 131072 
  \advance \auxa by 5
  \auxb=\ht\caixa
  \advance \auxb by \number\dp\caixa
  \divide \auxb by 131072 
  \advance \auxb by 5
  \ifcase \tipografo    
    \put{#4} at {\x} {\y}
  \or                   
    \ifcase #5 
      \put{#4} at {\x} {\y}
    \or        
      \put{\!pinta} at {\x} {\y}
      \advance \y by \number\auxb  
      \put{#4} at {\x} {\y}
    \or        
      \put{\!pinta} at {\x} {\y}
      \advance \auxa by -2  
      \advance \auxb by -2  
      \advance \x by \number\auxa  
      \advance \y by \number\auxb  
      \put{#4} at {\x} {\y}   
    \or        
      \put{\!pinta} at {\x} {\y}
      \advance \x by \number\auxa  
      \put{#4} at {\x} {\y}   
    \or        
      \put{\!pinta} at {\x} {\y}
      \advance \auxa by -2  
      \advance \auxb by -2  
      \advance \x by \number\auxa  
      \advance \y by -\number\auxb  
      \put{#4} at {\x} {\y}   
    \or        
      \put{\!pinta} at {\x} {\y}
      \advance \y by -\number\auxb  
      \put{#4} at {\x} {\y}   
    \or        
      \put{\!pinta} at {\x} {\y}
      \advance \auxa by -2  
      \advance \auxb by -2  
      \advance \x by -\number\auxa  
      \advance \y by -\number\auxb  
      \put{#4} at {\x} {\y}   
    \or        
      \put{\!pinta} at {\x} {\y}
      \advance \x by -\number\auxa  
      \put{#4} at {\x} {\y}   
    \or        
      \put{\!pinta} at {\x} {\y}
      \advance \auxa by -2  
      \advance \auxb by -2  
      \advance \x by -\number\auxa  
      \advance \y by \number\auxb  
      \put{#4} at {\x} {\y}   
    \fi
  \or                   
    \ifcase #5 
      \put{#4} at {\x} {\y}
    \or        
      \put{\!pinta} at {\x} {\y}
      \advance \y by \number\auxb  
      \put{#4} at {\x} {\y}
    \or        
      \put{\!pinta} at {\x} {\y}
      \advance \auxa by -2  
      \advance \auxb by -2  
      \advance \x by \number\auxa  
      \advance \y by \number\auxb  
      \put{#4} at {\x} {\y}   
    \or        
      \put{\!pinta} at {\x} {\y}
      \advance \x by \number\auxa  
      \put{#4} at {\x} {\y}   
    \or        
      \put{\!pinta} at {\x} {\y}
      \advance \auxa by -2  
      \advance \auxb by -2
      \advance \x by \number\auxa  
      \advance \y by -\number\auxb 
      \put{#4} at {\x} {\y}   
    \or        
      \put{\!pinta} at {\x} {\y}
      \advance \y by -\number\auxb 
      \put{#4} at {\x} {\y}   
    \or        
      \put{\!pinta} at {\x} {\y}
      \advance \auxa by -2  
      \advance \auxb by -2
      \advance \x by -\number\auxa 
      \advance \y by -\number\auxb 
      \put{#4} at {\x} {\y}   
    \or        
      \put{\!pinta} at {\x} {\y}
      \advance \x by -\number\auxa 
      \put{#4} at {\x} {\y}   
    \or        
      \put{\!pinta} at {\x} {\y}
      \advance \auxa by -2  
      \advance \auxb by -2
      \advance \x by -\number\auxa 
      \advance \y by \number\auxb  
      \put{#4} at {\x} {\y}   
    \fi
   \else 
     \ifnum\auxa<\auxb 
       \aux=\auxb
     \else
       \aux=\auxa
     \fi
     \ifdim\wd\caixa<1em
       \dimen99 = 1em
       \aux=\dimen99 \divide \aux by 131072 
       \advance \aux by 5
     \fi
     \advance\aux by -2 
     \multiply\aux by 2 %
     \ifnum\aux<30
       \put{\circle{\aux}} [Bl] at {\x} {\y}
     \else
       \multiply\auxa by 2
       \multiply\auxb by 2
       \put{\oval(\auxa,\auxb)} [Bl] at {\x} {\y}
     \fi
     \put{#4} at {\x} {\y}
   \fi   
}

\catcode`!=12 


\newif\ifpdf
\ifx\pdfoutput\undefined
\pdffalse 
\else
\pdfoutput=1 
\pdftrue
\fi

\input miniltx

\def\@ifpackageloaded#1{}

\ifpdf
\def\Gin@driver{pdftex.def}
\else
\def\Gin@driver{dvips.def}
\fi

\input graphicx.sty

\resetatcatcode

%
%

%
%
%
%

\def\Serif{cmr}
\def\SerifBold{cmbx}
\def\SerifItalics{cmti}
\def\SerifSlanted{cmsl}
\def\SerifBoldItalics{cmbxti}
\def\SansSerif{cmss}
\def\SansSerifBold{cmssbx}
\def\SansSerifItalics{cmssi}
\def\SansSerifSlanted{cmssi}
\def\Math{cmmi}
\def\Symbols{cmsy}
\def\MathBold{cmmib}
\def\MoreSymbols{cmex}
\def\Typewriter{cmtt}
\def\Gothic{eufm}
\def\Double{msbm}

= 			\Serif10 			at 5pt
= 		\SerifBold10 		at 5pt
= 	\SerifItalics10 	at 5pt
=		\SerifSlanted10 	at 5pt
=	\SerifBoldItalics10	at 5pt
= 		\SansSerif10 		at 5pt
=	\SansSerifBold10	at 5pt
=	\SansSerifItalics10	at 5pt
=	\SansSerifSlanted10	at 5pt
=				\Math10				at 5pt
=			\MathBold10			at 5pt
=			\Symbols10			at 5pt
=		\MoreSymbols10		at 5pt
=		\Typewriter10		at 5pt
=			\Gothic10			at 5pt
=			\Double10			at 5pt

= 			\Serif10 			at 7pt
= 		\SerifBold10 		at 7pt
= 	\SerifItalics10 	at 7pt
=	\SerifSlanted10 	at 7pt
=\SerifBoldItalics10	at 7pt
= 		\SansSerif10 		at 7pt
= 	\SansSerifBold10 	at 7pt
=\SansSerifItalics10	at 7pt
=\SansSerifSlanted10	at 7pt
=			\Math10				at 7pt
=		\MathBold10			at 7pt
=			\Symbols10			at 7pt
=		\MoreSymbols10		at 7pt
=		\Typewriter10		at 7pt
=			\Gothic10			at 7pt
=			\Double10			at 7pt

= 			\Serif10 			at 8pt
= 		\SerifBold10 		at 8pt
= 	\SerifItalics10 	at 8pt
=	\SerifSlanted10 	at 8pt
=\SerifBoldItalics10	at 8pt
= 		\SansSerif10 		at 8pt
= 	\SansSerifBold10 	at 8pt
=\SansSerifItalics10 at 8pt
=\SansSerifSlanted10 at 8pt
=			\Math10				at 8pt
=		\MathBold10			at 8pt
=			\Symbols10			at 8pt
=		\MoreSymbols10		at 8pt
=		\Typewriter10		at 8pt
=			\Gothic10			at 8pt
=			\Double10			at 8pt

= 			\Serif10 			at 10pt
= 		\SerifBold10 		at 10pt
= 		\SerifItalics10 	at 10pt
=		\SerifSlanted10 	at 10pt
=	\SerifBoldItalics10	at 10pt
= 		\SansSerif10 		at 10pt
= 	\SansSerifBold10 	at 10pt
= 	\SansSerifItalics10 at 10pt
= 	\SansSerifSlanted10 at 10pt
=				\Math10				at 10pt
=			\MathBold10			at 10pt
=			\Symbols10			at 10pt
=		\MoreSymbols10		at 10pt
=		\Typewriter10		at 10pt
=			\Gothic10			at 10pt
=			\Double10			at 10pt

= 				\Serif10 			at 12pt
= 			\SerifBold10 		at 12pt
= 		\SerifItalics10 	at 12pt
=		\SerifSlanted10 	at 12pt
=	\SerifBoldItalics10	at 12pt
= 			\SansSerif10 		at 12pt
= 		\SansSerifBold10 	at 12pt
= 	\SansSerifItalics10 at 12pt
= 	\SansSerifSlanted10 at 12pt
=				\Math10				at 12pt
=			\MathBold10			at 12pt
=			\Symbols10			at 12pt
=		\MoreSymbols10		at 12pt
=			\Typewriter10		at 12pt
=				\Gothic10			at 12pt
=				\Double10			at 12pt

= 			\Serif10 			at 14pt
= 		\SerifBold10 		at 14pt
= 	\SerifItalics10 	at 14pt
=		\SerifSlanted10 	at 14pt
=	\SerifBoldItalics10	at 14pt
= 		\SansSerif10 		at 14pt
= 	\SansSerifBold10 	at 14pt
= \SansSerifSlanted10 at 14pt
= \SansSerifItalics10 at 14pt
=				\Math10				at 14pt
=			\MathBold10			at 14pt
=			\Symbols10			at 14pt
=		\MoreSymbols10		at 14pt
=		\Typewriter10		at 14pt
=			\Gothic10			at 14pt
=			\Double10			at 14pt

\def\NormalStyle{\parindent=5pt\parskip=3pt\normalbaselineskip=14pt%
\def\nt{\tenSerif}%
\def\rm{\fam0\tenSerif}%
\textfont0=\tenSerif\scriptfont0=\sevenSerif\scriptscriptfont0=\fiveSerif
\textfont1=\tenMath\scriptfont1=\sevenMath\scriptscriptfont1=\fiveMath
\textfont2=\tenSymbols\scriptfont2=\sevenSymbols\scriptscriptfont2=\fiveSymbols
\textfont3=\tenMoreSymbols\scriptfont3=\sevenMoreSymbols\scriptscriptfont3=\fiveMoreSymbols
\textfont\itfam=\tenSerifItalics\def\it{\fam\itfam\tenSerifItalics}%
\textfont\slfam=\tenSerifSlanted\def\sl{\fam\slfam\tenSerifSlanted}%
\textfont\ttfam=\tenTypewriter\def\tt{\fam\ttfam\tenTypewriter}%
\textfont\bffam=\tenSerifBold%
\def\bf{\fam\bffam\tenSerifBold}\scriptfont\bffam=\sevenSerifBold\scriptscriptfont\bffam=\fiveSerifBold%
\def\cal{\tenSymbols}%
\def\greekbold{\tenMathBold}%
\def\gothic{\tenGothic}%
\def\Bbb{\tenDouble}%
\def\LieFont{\tenSerifItalics}%
\nt\normalbaselines\baselineskip=14pt%
}

\def\TitleStyle{\parindent=0pt\parskip=0pt\normalbaselineskip=15pt%
\def\nt{\fourteenSansSerifBold}%
\def\rm{\fam0\fourteenSansSerifBold}%
\textfont0=\fourteenSansSerifBold\scriptfont0=\tenSansSerifBold\scriptscriptfont0=\eightSansSerifBold
\textfont1=\fourteenMath\scriptfont1=\tenMath\scriptscriptfont1=\eightMath
\textfont2=\fourteenSymbols\scriptfont2=\tenSymbols\scriptscriptfont2=\eightSymbols
\textfont3=\fourteenMoreSymbols\scriptfont3=\tenMoreSymbols\scriptscriptfont3=\eightMoreSymbols
\textfont\itfam=\fourteenSansSerifItalics\def\it{\fam\itfam\fourteenSansSerifItalics}%
\textfont\slfam=\fourteenSansSerifSlanted\def\sl{\fam\slfam\fourteenSerifSansSlanted}%
\textfont\ttfam=\fourteenTypewriter\def\tt{\fam\ttfam\fourteenTypewriter}%
\textfont\bffam=\fourteenSansSerif%
\def\bf{\fam\bffam\fourteenSansSerif}\scriptfont\bffam=\tenSansSerif\scriptscriptfont\bffam=\eightSansSerif%
\def\cal{\fourteenSymbols}%
\def\greekbold{\fourteenMathBold}%
\def\gothic{\fourteenGothic}%
\def\Bbb{\fourteenDouble}%
\def\LieFont{\fourteenSerifItalics}%
\nt\normalbaselines\baselineskip=15pt%
}

\def\PartStyle{\parindent=0pt\parskip=0pt\normalbaselineskip=15pt%
\def\nt{\fourteenSansSerifBold}%
\def\rm{\fam0\fourteenSansSerifBold}%
\textfont0=\fourteenSansSerifBold\scriptfont0=\tenSansSerifBold\scriptscriptfont0=\eightSansSerifBold
\textfont1=\fourteenMath\scriptfont1=\tenMath\scriptscriptfont1=\eightMath
\textfont2=\fourteenSymbols\scriptfont2=\tenSymbols\scriptscriptfont2=\eightSymbols
\textfont3=\fourteenMoreSymbols\scriptfont3=\tenMoreSymbols\scriptscriptfont3=\eightMoreSymbols
\textfont\itfam=\fourteenSansSerifItalics\def\it{\fam\itfam\fourteenSansSerifItalics}%
\textfont\slfam=\fourteenSansSerifSlanted\def\sl{\fam\slfam\fourteenSerifSansSlanted}%
\textfont\ttfam=\fourteenTypewriter\def\tt{\fam\ttfam\fourteenTypewriter}%
\textfont\bffam=\fourteenSansSerif%
\def\bf{\fam\bffam\fourteenSansSerif}\scriptfont\bffam=\tenSansSerif\scriptscriptfont\bffam=\eightSansSerif%
\def\cal{\fourteenSymbols}%
\def\greekbold{\fourteenMathBold}%
\def\gothic{\fourteenGothic}%
\def\Bbb{\fourteenDouble}%
\def\LieFont{\fourteenSerifItalics}%
\nt\normalbaselines\baselineskip=15pt%
}

\def\ChapterStyle{\parindent=0pt\parskip=0pt\normalbaselineskip=15pt%
\def\nt{\fourteenSansSerifBold}%
\def\rm{\fam0\fourteenSansSerifBold}%
\textfont0=\fourteenSansSerifBold\scriptfont0=\tenSansSerifBold\scriptscriptfont0=\eightSansSerifBold
\textfont1=\fourteenMath\scriptfont1=\tenMath\scriptscriptfont1=\eightMath
\textfont2=\fourteenSymbols\scriptfont2=\tenSymbols\scriptscriptfont2=\eightSymbols
\textfont3=\fourteenMoreSymbols\scriptfont3=\tenMoreSymbols\scriptscriptfont3=\eightMoreSymbols
\textfont\itfam=\fourteenSansSerifItalics\def\it{\fam\itfam\fourteenSansSerifItalics}%
\textfont\slfam=\fourteenSansSerifSlanted\def\sl{\fam\slfam\fourteenSerifSansSlanted}%
\textfont\ttfam=\fourteenTypewriter\def\tt{\fam\ttfam\fourteenTypewriter}%
\textfont\bffam=\fourteenSansSerif%
\def\bf{\fam\bffam\fourteenSansSerif}\scriptfont\bffam=\tenSansSerif\scriptscriptfont\bffam=\eightSansSerif%
\def\cal{\fourteenSymbols}%
\def\greekbold{\fourteenMathBold}%
\def\gothic{\fourteenGothic}%
\def\Bbb{\fourteenDouble}%
\def\LieFont{\fourteenSerifItalics}%
\nt\normalbaselines\baselineskip=15pt%
}

\def\SectionStyle{\parindent=0pt\parskip=0pt\normalbaselineskip=13pt%
\def\nt{\twelveSansSerifBold}%
\def\rm{\fam0\twelveSansSerifBold}%
\textfont0=\twelveSansSerifBold\scriptfont0=\eightSansSerifBold\scriptscriptfont0=\eightSansSerifBold
\textfont1=\twelveMath\scriptfont1=\eightMath\scriptscriptfont1=\eightMath
\textfont2=\twelveSymbols\scriptfont2=\eightSymbols\scriptscriptfont2=\eightSymbols
\textfont3=\twelveMoreSymbols\scriptfont3=\eightMoreSymbols\scriptscriptfont3=\eightMoreSymbols
\textfont\itfam=\twelveSansSerifItalics\def\it{\fam\itfam\twelveSansSerifItalics}%
\textfont\slfam=\twelveSansSerifSlanted\def\sl{\fam\slfam\twelveSerifSansSlanted}%
\textfont\ttfam=\twelveTypewriter\def\tt{\fam\ttfam\twelveTypewriter}%
\textfont\bffam=\twelveSansSerif%
\def\bf{\fam\bffam\twelveSansSerif}\scriptfont\bffam=\eightSansSerif\scriptscriptfont\bffam=\eightSansSerif%
\def\cal{\twelveSymbols}%
\def\bg{\twelveMathBold}%
\def\gothic{\twelveGothic}%
\def\Bbb{\twelveDouble}%
\def\LieFont{\twelveSerifItalics}%
\nt\normalbaselines\baselineskip=13pt%
}

\def\SubSectionStyle{\parindent=0pt\parskip=0pt\normalbaselineskip=13pt%
\def\nt{\twelveSansSerifItalics}%
\def\rm{\fam0\twelveSansSerifItalics}%
\textfont0=\twelveSansSerifItalics\scriptfont0=\eightSansSerifItalics\scriptscriptfont0=\eightSansSerifItalics%
\textfont1=\twelveMath\scriptfont1=\eightMath\scriptscriptfont1=\eightMath%
\textfont2=\twelveSymbols\scriptfont2=\eightSymbols\scriptscriptfont2=\eightSymbols%
\textfont3=\twelveMoreSymbols\scriptfont3=\eightMoreSymbols\scriptscriptfont3=\eightMoreSymbols%
\textfont\itfam=\twelveSansSerif\def\it{\fam\itfam\twelveSansSerif}%
\textfont\slfam=\twelveSansSerifSlanted\def\sl{\fam\slfam\twelveSerifSansSlanted}%
\textfont\ttfam=\twelveTypewriter\def\tt{\fam\ttfam\twelveTypewriter}%
\textfont\bffam=\twelveSansSerifBold%
\def\bf{\fam\bffam\twelveSansSerifBold}\scriptfont\bffam=\eightSansSerifBold\scriptscriptfont\bffam=\eightSansSerifBold%
\def\cal{\twelveSymbols}%
\def\greekbold{\twelveMathBold}%
\def\gothic{\twelveGothic}%
\def\Bbb{\twelveDouble}%
\def\LieFont{\twelveSerifItalics}%
\nt\normalbaselines\baselineskip=13pt%
}

\def\AuthorStyle{\parindent=0pt\parskip=0pt\normalbaselineskip=14pt%
\def\nt{\tenSerif}%
\def\rm{\fam0\tenSerif}%
\textfont0=\tenSerif\scriptfont0=\sevenSerif\scriptscriptfont0=\fiveSerif
\textfont1=\tenMath\scriptfont1=\sevenMath\scriptscriptfont1=\fiveMath
\textfont2=\tenSymbols\scriptfont2=\sevenSymbols\scriptscriptfont2=\fiveSymbols
\textfont3=\tenMoreSymbols\scriptfont3=\sevenMoreSymbols\scriptscriptfont3=\fiveMoreSymbols
\textfont\itfam=\tenSerifItalics\def\it{\fam\itfam\tenSerifItalics}%
\textfont\slfam=\tenSerifSlanted\def\sl{\fam\slfam\tenSerifSlanted}%
\textfont\ttfam=\tenTypewriter\def\tt{\fam\ttfam\tenTypewriter}%
\textfont\bffam=\tenSerifBold%
\def\bf{\fam\bffam\tenSerifBold}\scriptfont\bffam=\sevenSerifBold\scriptscriptfont\bffam=\fiveSerifBold%
\def\cal{\tenSymbols}%
\def\greekbold{\tenMathBold}%
\def\gothic{\tenGothic}%
\def\Bbb{\tenDouble}%
\def\LieFont{\tenSerifItalics}%
\nt\normalbaselines\baselineskip=14pt%
}

\def\AddressStyle{\parindent=0pt\parskip=0pt\normalbaselineskip=14pt%
\def\nt{\eightSerif}%
\def\rm{\fam0\eightSerif}%
\textfont0=\eightSerif\scriptfont0=\sevenSerif\scriptscriptfont0=\fiveSerif
\textfont1=\eightMath\scriptfont1=\sevenMath\scriptscriptfont1=\fiveMath
\textfont2=\eightSymbols\scriptfont2=\sevenSymbols\scriptscriptfont2=\fiveSymbols
\textfont3=\eightMoreSymbols\scriptfont3=\sevenMoreSymbols\scriptscriptfont3=\fiveMoreSymbols
\textfont\itfam=\eightSerifItalics\def\it{\fam\itfam\eightSerifItalics}%
\textfont\slfam=\eightSerifSlanted\def\sl{\fam\slfam\eightSerifSlanted}%
\textfont\ttfam=\eightTypewriter\def\tt{\fam\ttfam\eightTypewriter}%
\textfont\bffam=\eightSerifBold%
\def\bf{\fam\bffam\eightSerifBold}\scriptfont\bffam=\sevenSerifBold\scriptscriptfont\bffam=\fiveSerifBold%
\def\cal{\eightSymbols}%
\def\greekbold{\eightMathBold}%
\def\gothic{\eightGothic}%
\def\Bbb{\eightDouble}%
\def\LieFont{\eightSerifItalics}%
\nt\normalbaselines\baselineskip=14pt%
}

\def\AbstractStyle{\parindent=0pt\parskip=0pt\normalbaselineskip=12pt%
\def\nt{\eightSerif}%
\def\rm{\fam0\eightSerif}%
\textfont0=\eightSerif\scriptfont0=\sevenSerif\scriptscriptfont0=\fiveSerif
\textfont1=\eightMath\scriptfont1=\sevenMath\scriptscriptfont1=\fiveMath
\textfont2=\eightSymbols\scriptfont2=\sevenSymbols\scriptscriptfont2=\fiveSymbols
\textfont3=\eightMoreSymbols\scriptfont3=\sevenMoreSymbols\scriptscriptfont3=\fiveMoreSymbols
\textfont\itfam=\eightSerifItalics\def\it{\fam\itfam\eightSerifItalics}%
\textfont\slfam=\eightSerifSlanted\def\sl{\fam\slfam\eightSerifSlanted}%
\textfont\ttfam=\eightTypewriter\def\tt{\fam\ttfam\eightTypewriter}%
\textfont\bffam=\eightSerifBold%
\def\bf{\fam\bffam\eightSerifBold}\scriptfont\bffam=\sevenSerifBold\scriptscriptfont\bffam=\fiveSerifBold%
\def\cal{\eightSymbols}%
\def\greekbold{\eightMathBold}%
\def\gothic{\eightGothic}%
\def\Bbb{\eightDouble}%
\def\LieFont{\eightSerifItalics}%
\nt\normalbaselines\baselineskip=12pt%
}

\def\RefsStyle{\parindent=0pt\parskip=0pt%
\def\nt{\eightSerif}%
\def\rm{\fam0\eightSerif}%
\textfont0=\eightSerif\scriptfont0=\sevenSerif\scriptscriptfont0=\fiveSerif
\textfont1=\eightMath\scriptfont1=\sevenMath\scriptscriptfont1=\fiveMath
\textfont2=\eightSymbols\scriptfont2=\sevenSymbols\scriptscriptfont2=\fiveSymbols
\textfont3=\eightMoreSymbols\scriptfont3=\sevenMoreSymbols\scriptscriptfont3=\fiveMoreSymbols
\textfont\itfam=\eightSerifItalics\def\it{\fam\itfam\eightSerifItalics}%
\textfont\slfam=\eightSerifSlanted\def\sl{\fam\slfam\eightSerifSlanted}%
\textfont\ttfam=\eightTypewriter\def\tt{\fam\ttfam\eightTypewriter}%
\textfont\bffam=\eightSerifBold%
\def\bf{\fam\bffam\eightSerifBold}\scriptfont\bffam=\sevenSerifBold\scriptscriptfont\bffam=\fiveSerifBold%
\def\cal{\eightSymbols}%
\def\greekbold{\eightMathBold}%
\def\gothic{\eightGothic}%
\def\Bbb{\eightDouble}%
\def\LieFont{\eightSerifItalics}%
\nt\normalbaselines\baselineskip=10pt%
}



%
%


\def\ModeYes{yes}
\def\ModeNo{no}

\def\ModeUndef{undefined}


\def\nx{\noexpand}
\def\ni{\noindent}
\def\newpage{\vfill\eject}

\def\ss{\vskip 5pt}
\def\ms{\vskip 10pt}
\def\bs{\vskip 20pt}

 \def\,{\mskip\thinmuskip}
 \def\!{\mskip-\thinmuskip}
 \def\>{\mskip\medmuskip}
 \def\;{\mskip\thickmuskip}

%
% ... 
%

\def\refsModePost{post}
\def\refsModeAuto{auto}

\def\dbRefsSatusModeOk{ok}
\def\dbRefsSatusModeError{error}
\def\dbRefsSatusModeWarning{warning}


\newcount\BNUM
\BNUM=0

\def\refs{}

\def\SetModePost{\xdef\refsMode{\refsModePost}}			
\SetModePost

\def\dbRefsStatusOk{%
	\xdef\dbRefsStatus{\dbRefsSatusModeOk}%
	\xdef\dbRefsError{\ModeNo}%
	\xdef\dbRefsWarning{\ModeNo}%
	\xdef\dbRefsInfo{\ModeNo}%
}

\def\dbRefs{%
}

\def\dbRefsGet#1{%
	\xdef\found{N}\xdef\ikey{#1}\dbRefsStatusOk%
	\xdef\key{\ModeUndef}\xdef\tag{\ModeUndef}\xdef\tail{\ModeUndef}%
	\dbRefs%
}

\def\NextRefsTag{%
	\global\advance\BNUM by 1%
}
\def\ShowTag#1{{\bf [#1]}}

\def\dbRefsInsert#1#2{%
\dbRefsGet{#1}%
\if\found Y %
   \xdef\dbRefsStatus{\dbRefsSatusModeWarning}%
   \xdef\dbRefsWarning{record is already there}%
   \xdef\dbRefsInfo{record not inserted}%
\else%
   \toks2=\expandafter{\dbRefs}%
   \ifx\refsMode\refsModeAuto \NextRefsTag
    \xdef\dbRefs{%
   	\the\toks2 \nx\xdef\nx\dbx{#1}%
	\nx\ifx\nx\ikey %
		\nx\dbx\nx\xdef\nx\found{Y}%
		\nx\xdef\nx\key{#1}%
		\nx\xdef\nx\tag{\the\BNUM}%
		\nx\xdef\nx\tail{#2}%
	\nx\fi}%
	\global\xdef\refs{\refs \ss\ni[\the\BNUM]\ #2\par}
   \fi%
   \ifx\refsMode\refsModePost 
    \xdef\dbRefs{%
   	\the\toks2 \nx\xdef\nx\dbx{#1}%
	\nx\ifx\nx\ikey %
		\nx\dbx\nx\xdef\nx\found{Y}%
		\nx\xdef\nx\key{#1}%
		\nx\xdef\nx\tag{\ModeUndef}%
		\nx\xdef\nx\tail{#2}%
	\nx\fi}%
   \fi%
\fi%
}

\def\dbRefsEdit#1#2#3{\dbRefsGet{#1}%
\if\found N 
   \xdef\dbRefsStatus{\dbRefsSatusModeError}%
   \xdef\dbRefsError{record is not there}%
   \xdef\dbRefsInfo{record not edited}%
\else%
   \toks2=\expandafter{\dbRefs}%
   \xdef\dbRefs{\the\toks2%
   \nx\xdef\nx\dbx{#1}%
   \nx\ifx\nx\ikey\nx\dbx %
	\nx\xdef\nx\found{Y}%
	\nx\xdef\nx\key{#1}%
	\nx\xdef\nx\tag{#2}%
	\nx\xdef\nx\tail{#3}%
   \nx\fi}%
\fi%
}

\def\bib#1#2{\RefsStyle\dbRefsInsert{#1}{#2}%
	\ifx\dbRefsStatus\dbRefsSatusModeWarning %
		\message{^^J}%
		\message{WARNING: Reference [#1] is doubled.^^J}%
	\fi%
}

\def\ref#1{\dbRefsGet{#1}%
\ifx\found N %
  \message{^^J}%
  \message{ERROR: Reference [#1] unknown.^^J}%
  \ShowTag{??}%
\else%
	\ifx\tag\ModeUndef \NextRefsTag%
		\dbRefsEdit{#1}{\the\BNUM}{\tail}%
		\dbRefsGet{#1}%
		\global\xdef\refs{\refs \ss\ni [\tag]\ \tail\par}
	\fi
	\ShowTag{\tag}%
\fi%
}

\def\ShowBiblio{\bs\Ensure{\SectionEnsure}%
{\SectionStyle\ni References}%
{\RefsStyle\refs}%
}

\newcount\CHANGES
\CHANGES=0
\def\AuxFile{7}
\def\PreventDoubleOn{\xdef\PreventDoubleLabel{\ModeYes}}

\PreventDoubleOn

\def\StoreLabel#1#2{\xdef\itag{#2}
 \ifx\PreModeStatus\ModeNo %
   \message{^^J}%
   \errmessage{You can't use Check without starting with OpenPreMode (and finishing with ClosePreMode)^^J}%
 \else%
   \immediate\write\AuxFile{\nx\dbLabelPreInsert{#1}{\itag}}%
   \dbLabelGet{#1}%
   \ifx\itag\tag %
   \else%
	\global\advance\CHANGES by 1%
 	\xdef\itag{(?.??)}%
    \fi%
   \fi%
}

\def\PreModeStatus{\ModeNo}

\def\ClosePreMode{\immediate\closeout\AuxFile%
  \ifnum\CHANGES=0%
	\message{^^J}%
	\message{**********************************^^J}%
	\message{**  NO CHANGES TO THE AuxFile  **^^J}%
	\message{**********************************^^J}%
 \else%
	\message{^^J}%
	\message{**************************************************^^J}%
	\message{**  PLAEASE TYPESET IT AGAIN (\the\CHANGES)  **^^J}%
    \errmessage{**************************************************^^ J}%
  \fi%
  \edef\PreModeStatus{\ModeNo}%
}

\def\dbLabelSatusModeOk{ok}

\def\dbLabelSatusModeWarning{warning}

\def\dbLabelStatusOk{%
	\xdef\dbLabelStatus{\dbLabelSatusModeOk}%
	\xdef\dbLabelError{\ModeNo}%
	\xdef\dbLabelWarning{\ModeNo}%
	\xdef\dbLabelInfo{\ModeNo}%
}

\def\dbLabel{%
}

\def\dbLabelGet#1{%
	\xdef\found{N}\xdef\ikey{#1}\dbLabelStatusOk%
	\xdef\key{\ModeUndef}\xdef\tag{\ModeUndef}\xdef\pre{\ModeUndef}%
	\dbLabel%
}

\def\ShowLabel#1{%
 \dbLabelGet{#1}%
 \ifx\tag \ModeUndef %
 	\global\advance\CHANGES by 1%
 	(?.??)%
 \else%
 	\tag%
 \fi%
}

\def\dbLabelPreInsert#1#2{\dbLabelGet{#1}%
\if\found Y %
  \xdef\dbLabelStatus{\dbLabelSatusModeWarning}%
   \xdef\dbLabelWarning{Label is already there}%
   \xdef\dbLabelInfo{Label not inserted}%
   \message{^^J}%
   \errmessage{Double pre definition of label [#1]^^J}%
\else%
   \toks2=\expandafter{\dbLabel}%
    \xdef\dbLabel{%
   	\the\toks2 \nx\xdef\nx\dbx{#1}%
	\nx\ifx\nx\ikey %
		\nx\dbx\nx\xdef\nx\found{Y}%
		\nx\xdef\nx\key{#1}%
		\nx\xdef\nx\tag{#2}%
		\nx\xdef\nx\pre{\ModeYes}%
	\nx\fi}%
\fi%
}

\def\dbLabelInsert#1#2{\dbLabelGet{#1}%
\xdef\itag{#2}%
\dbLabelGet{#1}%
\if\found Y %
	\ifx\tag\itag %
	\else%
	   \ifx\PreventDoubleLabel\ModeYes %
		\message{^^J}%
		\errmessage{Double definition of label [#1]^^J}%
	   \else%
		\message{^^J}%
		\message{Double definition of label [#1]^^J}%
	   \fi%
	\fi%
   \xdef\dbLabelStatus{\dbLabelSatusModeWarning}%
   \xdef\dbLabelWarning{Label is already there}%
   \xdef\dbLabelInfo{Label not inserted}%
\else%
   \toks2=\expandafter{\dbLabel}%
    \xdef\dbLabel{%
   	\the\toks2 \nx\xdef\nx\dbx{#1}%
	\nx\ifx\nx\ikey %
		\nx\dbx\nx\xdef\nx\found{Y}%
		\nx\xdef\nx\key{#1}%
		\nx\xdef\nx\tag{#2}%
		\nx\xdef\nx\pre{\ModeNo}%
	\nx\fi}%
\fi%
}


\newcount\PART
\newcount\CHAPTER
\newcount\SECTION
\newcount\SUBSECTION
\newcount\FNUMBER

\PART=0
\CHAPTER=0
\SECTION=0
\SUBSECTION=0	
\FNUMBER=0

\def\LastPart{\ModeUndef}
\def\LastChapter{\ModeUndef}
\def\LastSection{\ModeUndef}
\def\LastSubSection{\ModeUndef}
\def\LastClaim{\ModeUndef}
\def\Last{\ModeUndef}

\newdimen\TOBOTTOM
\newdimen\LIMIT

\def\Ensure#1{\ \par\ \immediate\LIMIT=#1\immediate\TOBOTTOM=\the\pagegoal\advance\TOBOTTOM by -\pagetotal%
\ifdim\TOBOTTOM<\LIMIT\newpage \else%
\vskip-\parskip\vskip-\parskip\vskip-\baselineskip\fi}

\def\PartLabel{\the\PART}
\def\NewPart#1{\global\advance\PART by 1%
         \bs\ni{\PartStyle  Part \PartLabel:}
         \bs\ni{\PartStyle #1}\newpage%
         \CHAPTER=0\SECTION=0\SUBSECTION=0\FNUMBER=0%
         \gdef\Left{#1}%
         \global\edef\Last{\PartLabel}%
         \global\edef\LastPart{\PartLabel}%
         \global\edef\LastChapter{\ModeUndef}%
         \global\edef\LastSection{\ModeUndef}%
         \global\edef\LastSubSection{\ModeUndef}%
         \global\edef\LastClaim{\ModeUndef}}
\def\ChapterLabel{\the\CHAPTER}
\def\NewChapter#1{\global\advance\CHAPTER by 1%
         \bs\ni{\ChapterStyle  Chapter \ChapterLabel: #1}\ms%
         \SECTION=0\SUBSECTION=0\FNUMBER=0%
         \gdef\Left{#1}%
         \global\edef\Last{\ChapterLabel}%
         \global\edef\LastChapter{\ChapterLabel}%
         \global\edef\LastSection{\ModeUndef}%
         \global\edef\LastSubSection{\ModeUndef}%
         \global\edef\LastClaim{\ModeUndef}}
\def\SectionEnsure{3cm}
\def\NewSection#1{\Ensure{\SectionEnsure}\gdef\SectionLabel{\the\SECTION}\global\advance\SECTION by 1%
         \bs\ni{\SectionStyle  \SectionLabel.\ #1}\ss%
         \SUBSECTION=0\FNUMBER=0%
         \gdef\Left{#1}%
         \global\edef\Last{\SectionLabel}%
         \global\edef\LastSection{\SectionLabel}%
         \global\edef\LastSubSection{\ModeUndef}%
         \global\edef\LastClaim{\ModeUndef}}
\def\NewAppendix#1#2{\Ensure{\SectionEnsure}\gdef\SectionLabel{#1}\global\advance\SECTION by 1%
         \bs\ni{\SectionStyle  Appendix \SectionLabel.\ #2}\ss%
         \SUBSECTION=0\FNUMBER=0%
         \gdef\Left{#2}%
         \global\edef\Last{\SectionLabel}%
         \global\edef\LastSection{\SectionLabel}%
         \global\edef\LastSubSection{\ModeUndef}%
         \global\edef\LastClaim{\ModeUndef}}
\def\Acknowledgements{\Ensure{\SectionEnsure}\gdef\SectionLabel{}%
         \bs\ni{\SectionStyle  Acknowledgments}\ss%
         \SECTION=0\SUBSECTION=0\FNUMBER=0%
         \gdef\Left{}%
         \global\edef\Last{\ModeUndef}%
         \global\edef\LastSection{\ModeUndef}%
         \global\edef\LastSubSection{\ModeUndef}%
         \global\edef\LastClaim{\ModeUndef}}
\def\SubSectionEnsure{2cm}
\def\SubSectionLabel{\ifnum\SECTION>0 \the\SECTION.\fi\the\SUBSECTION}
\def\NewSubSection#1{\Ensure{\SubSectionEnsure}\global\advance\SUBSECTION by 1%
         \ms\ni{\SubSectionStyle #1}\ss%
         \global\edef\Last{\SubSectionLabel}%
         \global\edef\LastSubSection{\SubSectionLabel}}
\def\SetNumberingModeN{\def\ClaimLabel{(\the\FNUMBER)}}
\def\SetNumberingModeSN{\def\ClaimLabel{(\ifnum\SECTION>0 \SectionLabel.\fi%
      \the\FNUMBER)}}
\def\SetNumberingModeCSN{\def\ClaimLabel{(\ifnum\CHAPTER>0 \the\CHAPTER.\fi%
      \ifnum\SECTION>0 \SectionLabel.\fi%
      \the\FNUMBER)}}

\def\NewClaim{\global\advance\FNUMBER by 1%
    \ClaimLabel%
    \global\edef\LastClaim{\ClaimLabel}%
    \global\edef\Last{\ClaimLabel}}

\def\HideLabels{\xdef\ShowLabelsMode{\ModeNo}}
\HideLabels

\def\fn{\eqno{\NewClaim}} 
\def\fl#1{%
\ifx\ShowLabelsMode\ModeYes%
 \eqno{{\buildrel{\hbox{\AbstractStyle[#1]}}\over{\hfill\NewClaim}}}%
\else%
 \eqno{\NewClaim}%
\fi%
\dbLabelInsert{#1}{\ClaimLabel}}
\def\fprel#1{\global\advance\FNUMBER by 1\StoreLabel{#1}{\ClaimLabel}%
\ifx\ShowLabelsMode\ModeYes%
\eqno{{\buildrel{\hbox{\AbstractStyle[#1]}}\over{\hfill.\itag}}}%
\else%
 \eqno{\itag}%
\fi%
}

\def\cl#1{\global\advance\FNUMBER by 1\dbLabelInsert{#1}{\ClaimLabel}%
\ifx\ShowLabelsMode\ModeYes%
${\buildrel{\hbox{\AbstractStyle[#1]}}\over{\hfill\ClaimLabel}}$%
\else%
  $\ClaimLabel$%
\fi%
}
\def\cprel#1{\global\advance\FNUMBER by 1\StoreLabel{#1}{\ClaimLabel}%
\ifx\ShowLabelsMode\ModeYes%
${\buildrel{\hbox{\AbstractStyle[#1]}}\over{\hfill.\itag}}$%
\else%
  $\itag$%
\fi%
}


\parindent=7pt
\leftskip=2cm
\newcount\SideIndent
\newcount\SideIndentTemp
\SideIndent=0
\newdimen\SectionIndent
\SectionIndent=-8pt

\def\sidebar{\vrule height15pt width.2pt }
\def\endcorner{\hbox{\hbox{\vrule height6pt width.2pt}\vbox to6pt{\vfill\hbox
to4pt{\leaders\hrule height0.2pt\hfill}}}}
\def\begincorner{\hbox{\hbox{\vrule height6pt width.2pt}\vbox to6pt{\hbox
to4pt{\leaders\hrule height0.2pt\hfill}}}}
\def\endbegincorner{\hbox{\vbox to15pt{\endcorner\vskip-6pt\begincorner\vfill}}}
\def\SideShow{\SideIndentTemp=\SideIndent \ifnum \SideIndentTemp>0 
\loop\sidebar\hskip 2pt \advance\SideIndentTemp by-1\ifnum \SideIndentTemp>1 \repeat\fi}

\def\BeginSection{{\vbadness 100000 \par\ni\hskip\SectionIndent%
\SideShow\vbox to 15pt{\vfill\begincorner}}\global\advance\SideIndent by1\vskip-10pt}

\def\EndSection{{\vbadness 100000 \par\ni\global\advance\SideIndent by-1%
\hskip\SectionIndent\SideShow\vbox to15pt{\endcorner\vfill}\vskip-10pt}}

\def\EndBeginSection{{\vbadness 100000\par\ni%
\global\advance\SideIndent by-1\hskip\SectionIndent\SideShow
\vbox to15pt{\vfill\endbegincorner}}%
\global\advance\SideIndent by1\vskip-10pt}

\def\ShowBeginCorners#1{%
\SideIndentTemp =#1 \advance\SideIndentTemp by-1%
\ifnum \SideIndentTemp>0 %
\vskip-15truept\hbox{\kern 2truept\vbox{\hbox{\begincorner}%
\ShowBeginCorners{\SideIndentTemp}\vskip-3truept}}%
\fi%
}

\def\ShowEndCorners#1{%
\SideIndentTemp =#1 \advance\SideIndentTemp by-1%
\ifnum \SideIndentTemp>0 %
\vskip-15truept\hbox{\kern 2truept\vbox{\hbox{\endcorner}%
\ShowEndCorners{\SideIndentTemp}\vskip 2truept}}%
\fi%
}

\def\BeginSections#1{{\vbadness 100000 \par\ni\hskip\SectionIndent%
\SideShow\vbox to 15pt{\vfill\ShowBeginCorners{#1}}}\global\advance\SideIndent by#1\vskip-10pt}

\def\EndSections#1{{\vbadness 100000 \par\ni\global\advance\SideIndent by-#1%
\hskip\SectionIndent\SideShow\vbox to15pt{\vskip15pt\ShowEndCorners{#1}\vfill}\vskip-10pt}}

\def\EndBeginSections#1#2{{\vbadness 100000\par\ni%
\global\advance\SideIndent by-#1%
\hbox{\hskip\SectionIndent\SideShow\kern-2pt%
\vbox to15pt{\vskip15pt\ShowEndCorners{#1}\vskip4pt\ShowBeginCorners{#2}}}}%
\global\advance\SideIndent by#2\vskip-10pt}




%
%


\def\al{\alpha}
\def\be{\beta}
\def\de{\delta}
\def\ga{\gamma}

\def\ep{\epsilon}

\def\te{\theta}
\def\la{\lambda}

\def\om{\omega}
\def\si{\sigma}
\def\vp{\varphi}

\def\ka{\kappa}

\def\Si{\Sigma}


 \def\su{{\hbox{\gothic su}}}

 \def\spin{{\hbox{\gothic spin}}}

 
 \def\A{{\hbox{\Bbb A}}}
 \def\R{{\hbox{\Bbb R}}}
 \def\C{{\hbox{\Bbb C}}}

 \def\R{{\hbox{\Bbb R}}}

 \def\K{{\hbox{\Bbb K}}}


\def\Aut{{\hbox{Aut}}}

\def\Ad{{\hbox{Ad}}}

\def\Spin{{\hbox{Spin}}}
\def\SO{{\hbox{SO}}}
\def\SU{{\hbox{SU}}}
\def\SL{{\hbox{SL}}}

\def\Diff{{\hbox{Diff}}}
\def\di{{\hbox{d}}}
\def\id{{\hbox{\rm id}}}

\def\ip{\hbox to4pt{\leaders\hrule height0.3pt\hfill}\vbox to8pt{\leaders\vrule width0.3pt\vfill}\kern 2pt}
 
\def\del{\partial}

\def\arr{\rightarrow}

%
%

\def\cases#1{\left\{\eqalign{#1}\right.}
\NormalStyle
\SetNumberingModeSN
\PreventDoubleOn

\long\def\title#1{\centerline{\TitleStyle\ni#1}}
\long\def\author#1{\ms\centerline{\AuthorStyle by {\it #1}}}

\long\def\address#1{\ss\centerline{\AddressStyle #1}\par}
\long\def\moreaddress#1{\centerline{\AddressStyle #1}\par}
\def\abstract{\ms\leftskip 3cm\rightskip .5cm\AbstractStyle{\bf \ni Abstract:}\ }
\def\endabstract{\par\leftskip 2cm\rightskip 0cm\NormalStyle\ss}

\SetNumberingModeSN

\def\frac[#1/#2]{\hbox{$#1\over#2$}}

\def\({\left(}
\def\){\right)}
\def\[{\left[}
\def\]{\right]}
\def\^#1{{}^{#1}_{\>\cdot}}
\def\_#1{{}_{#1}^{\>\cdot}}
\def\Label=#1{{\buildrel {\hbox{\fiveSerif \ShowLabel{#1}}}\over =}}
\def\<{\kern -1pt}

\bib{OurAshtekar}{L.\ Fatibene, M.\ Francaviglia, {\it Spin Structures on Manifolds and Ashtekar Variables},Int. J. Geom. Methods Mod. Phys. {\bf 2}(2) (2005), 147-157}

\bib{Book}{L.\ Fatibene, M.\ Francaviglia, {\it Natural and gauge natural formalism for classical field theories. A geometric perspective including spinors and gauge theories}, Kluwer Academic Publishers, Dordrecht, 2003}

\bib{Samuel}{J.\ Samuel, {\it Is Barbero's Hamiltonian Formulation a Gauge Theory of Lorentzian Gravity?}, Class.\ Quantum Grav.\ {\it 17}, 2000, 141-148}

\bib{Holst}{S.\ Holst, {\it Barbero's Hamiltonian Derived from a Generalized Hilbert-Palatini Action},
Phys.\ Rev.\ {\bf D53}, 5966, 1996}

\bib{RovelliBook}{C.\ Rovelli, {\it Quantum Gravity}, Cambridge University Press, Cambridge, 2004}

\bib{RovelliBI}{A.\ Perez, C.\ Rovelli, {\it Physical effects of the Immirzi parameter},
Phys.\ Rev.\ {\bf D73}, 044013, 2006}

\bib{gr-qc/0404018}{A.\ Ashtekar, J.\ Lewandowski,
{\it Background Independent Quantum Gravity: a Status Report}, gr-qc/0404018}

\bib{Thiemann}{T.\ Thiemann, {\it Introduction to Modern Canonical Quantum general Relativity};
gr-qc/0110034}

\bib{Barbero}{F.\ Barbero, {\it Real Ashtekar variables for Lorentzian signature space-time},
Phys.\ Rev.\ {\it D51}, 5507, 1996}

\bib{Immirzi}{G.\ Immirzi, {\it Quantum Gravity and Regge Calculus},
Nucl.\ Phys.\ Proc.\ Suppl.\ {\bf 57}, 65-72}

\bib{KobaNu}{S.\ Kobayashi, K.\ Nomizu,
  {\it Foundations of differential geometry},
  John Wiley \& Sons, Inc., New York, 1963 USA}

\bib{Antonsen}{F.\ Antonsen, M.S.N.\ Flagga, {\it Spacetime Topology (I) - Chirality and the Third Stiefel-Whitney Class}, Int.\ J.\ Th.\ Phys.\ {\bf 41}(2), 2002}

\bib{Milnor}{J.W.\ Milnor, J.\ Stasheff, {\it Characteristic classes}, Princeton University Press, 1974}

\bib{Bombelli}{L.Bombelli, 
{Lagrangian Formulation}, in: {\it New Perspectives in Canonical Gravity},
Eds. A.Ashtekar, Bibliopolis (Napoli, 1988)}

\bib{Thie2006}{T. Thiemann,
{\it LoopQuantumGravity: An InsideView}, hep-th/0608210}

\NormalStyle
\NormalStyle
\def\SelfDualEquivalenceAPP{A}

\title{On a Covariant Formulation of the Barbero-Immirzi Connection}

\author{L. Fatibene$^{1, 2}$, M. Francaviglia$^{1, 2, 3}$, C. Rovelli$^{4}$}

\address{$^1$ Department of Mathematics, University of Torino (Italy)}

\moreaddress{$^2$ INFN- Iniziativa Specifica Na12}

\moreaddress{$^3$ ESG - University of Cosenza (Italy)}

\moreaddress{${}^4$ Centre de Physique Th\'eorique de Luminy%
\footnote{$^\dagger$}{Unit\'e mixte de recherche (UMR 6207) du CNRS et des Universit\'es
de Provence (Aix-Marseille I), de la M\'editerran\'ee (Aix-Marseille
II) et du Sud (Toulon-Var); laboratoire affili\'e \`a la FRUMAM (FR
2291).},  Universit\'e de la M\'editerran\'ee, F-13288 Marseille, EU}

\abstract
The Barbero-Immirzi (BI) connection, as usually introduced out of a spin connection, is a global object though it does not transform properly as a genuine connection with respect to generic spin transformations, unless quite specific and suitable gauges are imposed.
We shall here investigate whether and under which global conditions a (properly transforming and {\it hence} global ) $\SU(2)$-connection can be canonically defined in a gauge covariant way.
Such $\SU(2)$-connection locally agrees with the usual BI connection and it can be defined on pretty general bundles; in particular triviality is not assumed.
As a by-product we shall also introduce a global covariant $\SU(2)$-connection over the whole spacetime (while for technical reasons  the BI connection in the standard formulation is just introduced on a space slice) which restricts to the usual BI connection on a space slice.
\endabstract

\NewSection{Introduction}

Loop Quantum Gravity (LQG) is usually introduced by starting from the selfdual formulation of GR; see \ref{RovelliBook}, \ref{gr-qc/0404018}, \ref{Thiemann} and references quoted therein.
In order to avoid complexification in the Lorentzian case (and the consequent reality constraints)  the framework goes through a (parametrical) canonical transformation in the Hamiltonian formulation defining a new one--parameter family of connections collectively called the {\it Barbero-Immirzi} (BI) connection. 
The Immirzi parameter which appears in the canonical transformation, and hence in the new variable field, does not affect vacuum field equations, while it appears in the equations when coupling to spinors (see \ref{RovelliBI}). 
It is hence a physical parameter which in LQG is later fixed by considering black hole entropy and assuming that the standard classical entropy, i.e.\ one quarter of the horizon area, is reproduced in the classical limit by microstate counting.
 
The BI connection is a generic $\SU(2)$-gauge connection on a $3d$ surface $S\subset M$ (where the Hamiltonian boundary conditions are set) in the $4d$ spacetime $M$. 
The BI connection so obtained is {\it global} since, by  some topological coincidences based on the groups and spacetime dimensions involved, the $\SU(2)$-principal bundles (below  denoted by ${}^+\<\Si$) over a (orientable, compact) $3d$ base $S$ on which BI connection is defined are necessarily trivial; see \ref{Thiemann} and references quoted therein. 

However,  the BI connection is constructed out of a generic spin connection defined on $M$.
The spin connection has its own transformation rules with respect to spin transformations and it induces 
spin transformations for the BI connection. Unfortunately, such induced transformations rules are {\it not}
in general the expected transformation rules for a good $\SU(2)$-connection.
In general, the induced transformation rules of the BI connection so obtained do not even factorize 
through an action of $\SU(2)$; they are and remain transformation rules just with respect to the spin group.

The situation is similar to the following simpler ``toy model'': let us consider $\R^2$ with a fixed global Cartesian coordinate system and two global {\it scalar} functions $(f, g)$.
It is obvious that, in {\it that} global coordinate system, the two functions do in fact define a global ``vector field'' $X(x)=f(x)\> i+g(x)\>j$ (where $(i,j)$ are the natural basis of tangent vectors induced by coordinates).
However, since $(f,g)$ are scalars, the ``vector field'' $X$ depends on the coordinates chosen. In a different coordinate system, in fact, the same functions define a different ``vector field''. Or equivalently, the transformation rules of the scalar functions induce transformations rules for the {\it ``vector field''} which are not the correct ones for the components of  a vector field. 

In the literature concerning LQG, once the BI connection is introduced and recognized as a global object on the $\SU(2)$-bundle ${}^+\<\Si$, canonical quantization is developed for such a $\SU(2)$-reduced theory.
Depending on the approach, the transformation rules are not considered (i.e.~one is concerned with the local formalism) or
the object is {\it promoted} to be a good $\SU(2)$-connection.
At least, a particular gauge fixing is understood to provide a drastic simplification of transition functions.
In either cases, the relation between gauge transformations of the reduced BI model and the original spin model is lost (or at least well hidden) forever.

We shall show hereafter that this current situation can be considerably improved under many viewpoints by simply writing  in the appropriate bundles the objects involved in the constructions.
This setting will enhance a better control on the covariance issues and the BI connection will be defined as a manifestly ``good'' $\SU(2)$-connection.
As a consequence of the improved transformation rules we shall be able to go through the BI construction not only on a given space slice (i.e.~on the bundle ${}^+\<\Si$ where  Hamiltonian framework is set and hence where BI connection is usually defined) but also on spacetime (i.e.~on the bundle which will be below denoted by ${}^+\<\<P$).
The reduced BI connection on ${}^+\<\Si$ will be hence defined as the restriction of the spacetime counterpart of the BI connection defined on ${}^+\<\<P$ which will in turn provide a ``spacetime interpretation'' of the reduced BI connection on ${}^+\<\Si$.

The price which we need to pay for this improvement is a new formulation of GR based on a new structure bundle;
the new formulation is locally {\it identical} to the old one, though differences are hidden in the global structure of the group of spin transformations and in some globality issues.
For this reason we shall not discuss hereafter dynamics, which is locally unchanged 
with respect to the usual framework. Holst's action (see \ref{Holst}) has exactly the same expression, though written on the new bundle (denoted below by $\hat P$). It induces the same field equations and the Hamiltonian formalism is performed in the same way.
The final output is a model for the BI connection on ${}^+\<\Si$ with the same field equations to be implemented. The only differences with respect to the standard formulation are that the new BI connection is now {\it by construction} a good $\SU(2)$-connection and no gauge fixing is needed;
in fact, the construction is manifestly gauge covariant.
As noted in \ref{RovelliBook}, LQG is the quantization of these equations; thus no news in the quantization procedure either.

We shall hereafter consider the Euclidean case first. In the Euclidean sector the BI connection is of course not needed, since the selfdual connection is already real.
However, the BI connection can be defined as in the Lorentzian case and it is affected by similar problems.

\NewSection{Notation and Selfdual Formulation}

We shall here review the standard setting for the selfdual formulation. This will fix notation for later discussion.

The selfdual formulation relies on a Lie algebra duality defined on $\spin(4)$ which allows us to split it as the product of two copies of the algebra $\su(2)$. Since we are going to discuss gauge connections (which are defined on principal bundles) we need to go deeper into the duality and start from the duality at group level. By a well-known fact in group theory, we have the following natural group splitting:
$$
p:\Spin(4)\arr \SU(2)\times\SU(2)
\qquad\qquad
p_\pm:\Spin(4)\arr \SU(2)
\fl{GroupDuality}$$
The group projections $p_\pm$ are defined so that $p(S)=(p_+(S), p_-(S))$. 
We shall systematically use this canonical group isomorphism to identify  an element in $\Spin(4)$ with a pair of   elements in $\SU(2)$.
Such morphism does extend to the algebra and induces by projection on the first (second) factor the splitting in the (anti)selfdual part of the elements of the algebra  $\spin(4)$ 
$$
T_e p_\pm:\spin(4)\arr \su(2)
\fn$$
These projections trivially extend to objects valued in the Lie algebras, such as the connections.
Let now $P$ be a $\Spin(4)$-principal bundle over a $m=4$ dimensional manifold $M$. Once a local trivialization (also known as a {\it gauge}) of $P$ is choosen, a point in $P$ is locally denoted by $(x, S_+, S_-)$ (with $x\in M$ and $S_+, S_-\in \SU(2)$). 
The group of gauge transformations (or changes of gauge fixing, depending on the active or passive viewpoint) is implemented as the group of principal automorphisms $\Aut(P)$ which locally read as
$$
\cases{
&x'= f(x)\cr
&(S'_+,S'_-) = \(\phi_+(x) \cdot S_+, \phi_-(x) \cdot S_-\)   \cr
}
\qquad\qquad
\phi_+(x), \phi_-(x)\in \SU(2)
\fn$$
Notice that gauge transformations  in $\Aut(P)$ project over spacetime diffeomorphisms in $\Diff(M)$ (though of course there is no global gauge invariant embedding of spacetime diffeomorphisms in $\Aut(P)$).
However, vertical gauge transformations in $\Aut_V(P)$, namely gauge transformations projecting on spacetime identity (i.e.~$x'=x$), are globally defined and they will be called {\it pure gauge transformations}.

Let us fix a basis of vertical right invariant vector fields $\si_{ab}$ on $P$; a connection on $P$ is locally represented as
$$
\eqalign{
&\om= \di x^\mu\otimes \( \del_\mu -\om^{ab}_\mu(x) \si_{ab}\) \cr
&\quad\hbox{or equivalently}
\>
\tilde\om= \({\bar \Ad}^{ab}_{\>cd}\> \om^{cd}_\mu(x) \di x^\mu + \te^{ab}_L\)\otimes T_{ab}
}
\fn$$
where $T_{ab}$ is a basis of the Lie algebra $\spin(4)$, $\te_L=\te^{ab}_L\otimes T_{ab}$ is the corresponding basis of left invariant $1$-forms on $P$ with values in the Lie algebra $\spin(4)$ (also known as the Maurer-Cartan form).
We refer to \ref{Book} for the global intrinsic notation; hereafter we shall be concerned with the connection coefficients $\om^{ab}_\mu(x) $.
Let us now consider a gauge transformation $\Phi\in\Aut(P)$ locally expressed as $(x, \si)\mapsto (f(x), \phi(x)\cdot \si)$ with $\phi(x)=(\phi_+(x), \phi_-(x))\in \Spin(4)$; the connection coefficients transform as
$$
\om'{}^{ab}_\mu=\bar J^\nu_\mu \ell^a_c(\phi)\>\(\om^{cd}_\nu\> \ell^b_d(\phi)+ \di_\nu \ell^c_d(\phi^{-1})\eta^{db}\)
\fl{SpinConnectionTranformationRules}$$
where $\bar J^\nu_\mu$ denotes the inverse Jacobian of the spacetime diffeomorphism $f$ and 
$\ell: \Spin(4) \arr \SO(4)$ is the covering map exhibiting the spin group as a double covering of the corresponding orthogonal group. Here $\eta^{db}$ denotes the (inverse) Minkowski metric of the relevant signature; since in this case the signature is $(4,0)$ then $\eta^{db}\equiv\de^{db}$.

We stress that, from a local viewpoint, when one defines global connections the transformation rules \ShowLabel{SpinConnectionTranformationRules} are in fact a relevant part of the definition. 

For later convenience, notice that the group $\SU(2)$ is in fact canonically isomorphic to $\Spin(3)$ and let us denote its covering map by $\la:\SU(2)\arr \SO(3)$.

Standard GR formalism deals with a {\it spin connection} $\om^{ab}_\mu$ on a given $\Spin(4)$-principal bundle $P$ over the spacetime $M$ and a {\it frame field} $e_a$ (also called a {\it tetrad}). Here we shall forget about the frame (for which we refer to \ref{RovelliBook}, \ref{OurAshtekar}) and pay attention to the connection only.
Let us here just remind for completeness that, as shown in \ref{OurAshtekar}, the role of tetrads is essential to define Hamiltonian framework avoiding gauge fixings (or {\it a priori} fixing a foliation) which would spoil manifest gauge covariance. 
In fact, the selfdual formulation of GR leaves the antiselfdual part of the connection undetermined;
one can use antiselfdual gauge freedom (which is irrelevat to the framework) to ``adapt the frame'' 
to the Cauchy surface $S\subset M$.
Equivalently, this is acheived without any gauge fixing once  ``frames'' are properly regarded as gauge natural objects; see \ref{OurAshtekar} for details.

When the selfdual connection is introduced one should first of all  define a $\SU(2)$-principal bundle out of $P$. This can be easily done by using the group homomorphism \ShowLabel{GroupDuality}.
The bundle $P$ is characterized by its cocycle of transition functions 
$\psi_{\la\be}:U_{\al\be}\arr\Spin(4)$; 
this cocycle can be projected onto a $\SU(2)$ cocycle
$p_+\circ \psi_{\la\be}:U_{\al\be}\arr\SU(2)$ which in turn defines a unique (up to bundle isomorphisms) $\SU(2)$-principal bundle ${}^+\<\<P$.  We refer to \ref{Book} for details.
The situation can be summarized by the following commutative diagram:
$$
\begindc{\commdiag}[1]
\obj(40,80)[P1]{$P$}
\obj(110,80)[+P2]{$^+\<\<P$}
\obj(40,30)[M1]{$M$}
\obj(110,30)[M2]{$M$}
\mor{P1}{M1}{}
\mor{+P2}{M2}{}
\mor{P1}{+P2}{$p_+$}
\mor{M1}{M2}{}[\atleft, \solidline] \mor(40,33)(110,33){}[\atleft, \solidline]
\enddc
\qquad
\begindc{\commdiag}[1]
\obj(70,80)[P1]{$(x, S_+, S_-)$}
\obj(150,80)[+P2]{$(x, S_+)$}
\obj(30,30)[ghost]{}
\mor{P1}{+P2}{$p_+$}
\enddc
\fn$$

Now the standard definition of the selfdual connection
$$
{}^+\<\om^k_\mu=  \frac[1/2]\ep^k{}_{ij}\> \om^{ij}_\mu+ \om^{0k}_\mu 
\fl{SelfdualConnection}$$
does in fact define a connection of the bundle ${}^+\<\<P$; in particular, the transformation rules 
of $\om$ induce transformation rules for ${}^+\<\<\om$ which read as (see Appendix 
$\SelfDualEquivalenceAPP$)
$$
{}^+\<\om'{}^{k}_\mu=\bar J^\nu_\mu \>\( \la^k_i(\phi_+){}^+\<\<\om^{i}_\nu
-\frac[1/2]\ep^{kj}{}_{i}  \la^i_l(\phi_+)\di_\nu \bar\la^l_j(\phi_+)\)
\fl{SelfdualConnectionTransformationRules}$$
We stress that the transformation rules \ShowLabel{SelfdualConnectionTransformationRules} emerge by very special algebraic facts which heavily rely on the specific form \ShowLabel{SelfdualConnection}
of the selfdual connection. In particular, the transformation rules of ${}^+\<\om^{k}_\mu$ just depend on the selfdual $\SU(2)$ part of the spin group.
In other words, despite it is defined in terms of a $\Spin(4)$ field $\om^{ab}_\mu$, the selfdual connection 
${}^+\<\om{}^{k}_\mu$ is in fact a $\SU(2)$-field.

In the Lorentzian sector one is forced to use a complexification of the spin group ---in order to keep resorting to the group isomorphism \ShowLabel{GroupDuality} though in its complexified form--- and consequently is forced to complexification of the  bundles and of the coefficients of the selfdual connection which are defined by
$$
{}^+\<\<\om^k_\mu= \frac[1/2]\ep^k{}_{ij}\> \om^{ij}_\mu+i \om^{0k}_\mu
\fl{LorentzianSelfdualConnection}$$

In either signature, once the selfdual connection is defined, the action (see \ref{Bombelli})
$$
L_+(e, j^1{}^+\om)= {}^+\<R^{ab}\land e^c \land e^d \>\ep_{abcd}=  p_i^{ab} F^{i}_{\mu\nu} e_a^\mu e_b^\nu \> \sqrt{g}\> \di s
\fn$$
is considered as a  replacement of  the standard Hilbert-Einstein action. This action just depends on the selfdual connection (the selfdual part of the curvature ${}^+\<R^{ab}$ coincides in fact with the curvature  $F^{i}$ of the selfdual connection). 
The corresponding field equations are equivalent to the standard GR equation; see \ref{RovelliBook}, \ref{gr-qc/0404018} and references quoted therein. 

The Hamiltonian formulation goes through the choice of an embedded Cauchy surface $i:S\arr M$; 
let us then choose a coordinate system $k^A$ on $S$ so that the canonical inclusion is locally given by functions $x(k)$.
By means of a standard construction one can pull-back (i.e.\ {\it restrict}) bundles on $M$ to $S\subset M$ obtaining the following situation:
$$
\begindc{\commdiag}[1]
\obj(40,80)[P1]{$P$}
\obj(110,80)[+P2]{$^+P$}
\obj(10,50)[Si3]{$\Si$}
\obj(70,50)[+Si4]{$^+\Si$}
\obj(40,30)[M1]{$M$}
\obj(110,30)[M2]{$M$}
\obj(10,0)[S3]{$S$}
\obj(70,0)[S4]{$S$}
\mor{P1}{M1}{}
\mor{+P2}{M2}{}
\mor{Si3}{S3}{}
\mor{+Si4}{S4}{}
\mor{P1}{+P2}{$p_+$}
\mor{M1}{M2}{}[\atleft, \solidline] \mor(40,33)(110,33){}[\atleft, \solidline]
\mor{S3}{S4}{}[\atleft, \solidline] \mor(10,3)(70,3){}[\atleft, \solidline]
\mor{S3}{M1}{}[\atleft, \injectionarrow]
\mor{S4}{M2}{}[\atleft, \injectionarrow]
\mor{Si3}{P1}{}[\atleft, \injectionarrow]
\mor{+Si4}{+P2}{}[\atleft, \injectionarrow]
\mor{Si3}{+Si4}{$\pi_+$}
\enddc
\qquad
\begindc{\commdiag}[1]
\obj(70,150)[P1]{$(x(k), S_+, S_-)$}
\obj(180,150)[+P2]{$(x(k), S_+)$}
\obj(20,100)[Si3]{$(k, S_+, S_-)$}
\obj(120,100)[+Si4]{$(k, S_+)$}
\obj(70,70)[ghost]{}
\mor{P1}{+P2}{$p_+$}
\mor{Si3}{P1}{}[\atleft, \injectionarrow]
\mor{+Si4}{+P2}{}[\atleft, \injectionarrow]
\mor{Si3}{+Si4}{$\pi_+$}
\enddc
\fn$$
The new bundles $\Si$ and ${}^+\<\Si$ are principal bundles over $S$ with structure groups $\Spin(4)$
and $\SU(2)$, respectively. We refer to \ref{Book} for technicalities.

The selfdual connection ${}^+\<\<\om^k_\mu$ on ${}^+\<\<P$ canonically induces by restriction
a selfdual connection ${}^+\<\<\om^k_A(k)= {}^+\<\<\om^k_\mu(x(k))\>\del_Ax^\mu(k)$ on ${}^+\<\Si$.
This restricted selfdual connection ${}^+\<\<\om^k_A(k)$ (together with the densitised triad field $E_k^A(k)$ induced by the frame; see \ref{RovelliBook}, \ref{OurAshtekar}) are canonical variables for the Hamiltonian formalism which is the starting point for LQG.

\NewSection{Barbero-Immirzi Connection}

In this Section we shall review the standard setting for the Barbero-Immirzi connection. 
Again this is for notation fixing. Global issues and the new framework for BI connection are postponed to the following Sections.

In order to avoid complexification in the Lorentzian case a parametrical canonical transformation was introduced in the Hamiltonian formulation. The transformation defines new real variables:
$$
A^k_A :=  \frac[1/2]\ep^k{}_{ij}\> \om^{ij}_A +\ga\om^{0k}_A
\qquad\quad\ga\in \R-\{0\}
\fl{BIConnection}$$
namely the BI connection. Here we set  $\om^{ab}_A=\om^{ab}_\mu \>\del_Ax^\mu$ for the objects restricted to $S$.
Let us also define the field
$$
K^k_A :=  \om^{0k}_A
\fl{BIK}$$
Notice that definitions \ShowLabel{BIConnection} and \ShowLabel{BIK} are given on $S\subset M$;
if one had to repeat the construction on $M$ we stress that ``canonical transformations''
are {\it undefined} at spacetime level and in general the bundle ${}^+\<\<P$ does not need to be trivial so that one would need to discuss transformation rules and/or gauge fixings to restore manifest gauge covariance.

Notice also that for $\ga=\pm1$ the BI connection reduces to the (anti)selfdual (Euclidean) connection.
The case $\ga=\pm1$ is degenerate in many algebraic viewpoints and it is usually considered separately. Hereafter $\ga$ is then assumed to have a real value different from $\ga\not=0, \pm1$.

Barbero (see \ref{Barbero}) proposed new variables for the Lorentzian case, obtained by setting $\ga=1$ in \ShowLabel{BIConnection} while Immirzi later extended the definition of \ref{Barbero} to a whole one-parameter family of connections (see \ref{Immirzi}).

The quantities $A^k_A $ define a global object on ${}^+\<\Si$ which is necessarily trivial.
However, the original connection $\om^{ab}_\mu$ has its own transformation rules
\ShowLabel{SpinConnectionTranformationRules} which in turn prescribe transformation rules of the
objects $(A^k_A, K^k_A)$.
One obtains:
$$
\eqalign{
A'{}^i_A =& \bar J^B_A\Big[
\(\frac[1/2]\ep^i{}_{jk} \ell^j_m  \ell^k_l \ep^{ml}{}_h
+\ga \ell^0_m\ell^i_l\ep^{ml}{}_h\) A^h_B+\cr
&+\frac[1/2]\ep^i{}_j{}^k \ell^j_0 \di_B \bar \ell^0_k
+\frac[1/2]\ep^i{}_j{}^k \ell^j_m \di_B \bar \ell^m_k
+\ga \ell^0_m\di_B \bar\ell^m_j\eta^{ji}
+\ga \ell^0_0\di_B \bar\ell^0_j\eta^{ji}+\cr
&+\ep^i{}_{jk} \ell^j_0\ell ^k_m K^m_B
+\ga\(\ell^0_0 \ell ^i_h
-\ell^0_h\ell^i_0
-\frac[1/2]\ep^i{}_{jk} \ell^j_m\ell ^k_l\ep^{ml}{}_h
\)K^h_B+\cr
&-\ga^2\ep^{mj}{}_h \ell^0_m\ell^i_j K^h_B
\Big]\cr
K'{}^i_A=& \bar J^B_A\Big[
\(\ell^0_0\ell^i_k-\ell^0_k\ell^i_0\)K^k_B
-\ga \ell^0_k\ell^i_j\ep^{kj}{}_l K^l_B
+ \ell^0_k\ell^i_j\ep^{kj}{}_l A^l_B+\cr
&+\(\ell^0_0\di_B \bar\ell^0_j +\ell^0_k\di_B \bar\ell^k_j\)\eta^{ji}
\Big]\cr
}
\fl{BadTransformationRules}$$
which are definitely different from the transformation rules of a connection unless (as we shall see in next Section) a very specific form for the gauge transformation is assumed.
We stress that $\ell^0_0$, $\ell^i_0$, $\ell^0_i$, $\ell^i_j$ denote the blocks of $\ell^a_b\in \SO(4)$
and hence no specific form can be assumed in general.
One can try with some explicit generic element of $\Spin(4)$ to show that extra terms in 
\ShowLabel{BadTransformationRules} do not vanish in general.
Because of this, one cannot assume $A^k_A$ (as a function of $\om^{ab}_A$) to be a global $\SU(2)$-connection, even considering the triviality of ${}^+\<\Si$.

On the other hand, as we shall see in the next Sections, one has to be careful in considering gauge transformations of a special form; globality of such special transformations needs to be carefully discussed since it might in principle impose topological restrictions on $M$ and $P$.

The dynamics is then described by the following Holst action (see \ref{gr-qc/0404018}, \ref{Holst} and references quoted therein):
$$
L_\ga= \frac[1/4\ka]R^{ab} \land e^c\land e^d \ep_{abcd} 
+\frac[1/2\ka\ga]R^{ab} \land e_a \land e_b
\fn$$
which induces Lagrangian field equations.  Then the theory is recasted into Hamiltonian form written in terms of the new variables $(A^k_A, E_k^A)$ and canonical quantization can be started.

\NewSection{Covariance of the BI Connection}

Until now we just reviewed the standard setting. Now we shall investigate the covariance properties of the BI connection.

The new setting relies on a nice algebraic fact: {\it if we  could} restrict the spin group to
the subgroup $\si=(S_+, S_+)$ (i.e.~the diagonal form with respect to the (anti)selfdual decomposition)
then we easily could  prove that
$$
\ell(\si)=\(\matrix{
1 & 0 \cr
0 & \la(S_+)\cr
}\)
\fl{SU2SpecialForm}$$
Notice that the subgroup $\si=(S_+, S_+)$ is an isomorphic embedding of $\SU(2)$ within $\Spin(4)$.

For elements in this simpler form the extra terms in \ShowLabel{BadTransformationRules}
do in fact vanish and the transformation rules obtained are the appropriate ones for a $\SU(2)$-connection:
$$
\eqalign{
A'{}^i_A =& \bar J^B_A\Big[
\frac[1/2]\ep^i{}_{jk} \la^j_m  \la^k_l \ep^{mn}{}_h A^h_B
+\frac[1/2]\ep^i{}_j{}^l \la^j_m \di_B \bar \la^m_l
+\ga\(\la^i_h
-\frac[1/2]\ep^i{}_{jk} \la^j_m\la^k_l\ep^{ml}{}_h
\)K^h_B\Big]=\cr
=&\bar J^B_A\Big[
\la^i_j  A^j_B
+\frac[1/2]\ep^i{}_j{}^l \la^j_m \di_B \bar \la^m_l
+\ga\(\la^i_h
-\la^i_h
\)K^h_B\Big]=\cr
=&\bar J^B_A\Big[
\la^i_j  A^j_B
+\frac[1/2]\ep^i{}_j{}^l \la^j_m \di_B \bar \la^m_l\Big]\cr
&\cr
K'{}^i_A=& \bar J^B_A \la^i_j  K^j_B\cr
}
\fl{GoodTransformationRulesBI}$$

Hence we should only investigate when under and which conditions one is allowed to consider the subgroup of gauge transformations in the form $\si=(S, S)$. The issue is not trivial since the local expression for a $\Spin(4)$-gauge transformation $\phi$ as a pair of $\SU(2)$-gauge transformations $\phi=(\phi_+, \phi_-)$ does in fact depend on the trivialization chosen on $P$.
Even tuning $\phi_+=\phi_-$ in a given trivialization this form has no intrinsic meaning; 
when the trivialization is changed the special form is not preserved in general.

In fact, transition functions of $P$ are in general of the form $(\vp_+, \vp_-)$ so that in the new trivialization the same gauge transformation is generated by
$(\vp_+\cdot\phi_+, \vp_-\cdot\phi_+)$ which is not in the special form any longer.

The only case in which the special subgroup is intrinsic is when $P$ has some special trivialization
with transition functions in the special form $\vp_+=\vp_-$.
When this happens one says that $P$ admits a reduction from the group $\Spin(4)$ to the group $\SU(2)$, or in short a $\SU(2)$-reduction; see \ref{KobaNu}. 
This corresponds to ask that one can cover the whole spacetime with patches choosing a local gauge in each patch such that all transition functions among different local gauges are in the special form $(\vp_+, \vp_+)$. 

Of course one could {\it assume} $P$ to have such $\SU(2)$-reduction, which usually restricts the allowed $P$ and possibly imposes topological restrictions on $M$.
We shall show hereafter that one can always explicitly define out of $P$ a new bundle $\hat P$ having such a $\SU(2)$-reduction. 

We shall not investigate here in the Euclidean case whether the new bundle $\hat P$ is in fact different (i.e.~non-isomorphic) from $P$. In the Euclidean sector we are considerably less confident than in the Lorentzian case about which topological restrictions are ``physically reasonable'' for the ``spacetime'' $M$.
For this reason we choose to assume as less as possible about $M$ and $P$.

The canonical prescription for defining $\hat P$ again relies on a group homorphism
$$
\begindc{\commdiag}[1]
\obj(10,10)[Spin]{$\Spin(4)$}
\obj(60,10)[SU]{$\SU(2)$}
\obj(110,10)[SUxSU]{$\Spin(4)$}
\mor{Spin}{SU}{}
\mor{SU}{SUxSU}{$\be$}
\obj(10,30)[spin]{$(S_+, S_-)$}
\obj(60,30)[sU]{$S_+$}
\obj(110,30)[sUxSU]{$(S_+, S_+)$}
\mor{spin}{sU}{}
\mor{sU}{sUxSU}{}
\enddc
\fn$$
Using the group homomorphism $\be$ we can map the transition functions of ${}^+\<\<P$
to define a new cocycle $\be\circ\vp_+$ with values in $\Spin(4)$.
Using this new cocycle we can define (uniquely up to isomorphisms) a new bundle $\hat P$.
The construction is summarized by the following diagram
$$
\begindc{\commdiag}[1]
\obj(40,80)[P1]{$P$}
\obj(110,80)[+P2]{$^+P$}
\obj(180,80)[hP3]{$\hat P$}
\obj(10,50)[Si3]{$\Si$}
\obj(70,50)[+Si4]{$^+\Si$}
\obj(140,50)[hSi5]{$\hat\Si$}
\obj(40,30)[M1]{$M$}
\obj(110,30)[M2]{$M$}
\obj(180,30)[M3]{$M$}
\obj(10,0)[S3]{$S$}
\obj(70,0)[S4]{$S$}
\obj(140,0)[S5]{$S$}
\mor{P1}{M1}{}
\mor{+P2}{M2}{}
\mor{hP3}{M3}{}
\mor{Si3}{S3}{}
\mor{+Si4}{S4}{}
\mor{hSi5}{S5}{}
\mor{P1}{+P2}{$p_+$}
\mor{+P2}{hP3}{}
\mor{M1}{M2}{}[\atleft, \solidline] \mor(40,33)(110,33){}[\atleft, \solidline]
\mor{M2}{M3}{}[\atleft, \solidline] \mor(110,33)(180,33){}[\atleft, \solidline]
\mor{S3}{S4}{}[\atleft, \solidline] \mor(10,3)(70,3){}[\atleft, \solidline]
\mor{S4}{S5}{}[\atleft, \solidline] \mor(70,3)(140,3){}[\atleft, \solidline]
\mor{S3}{M1}{}[\atleft, \injectionarrow]
\mor{S4}{M2}{}[\atleft, \injectionarrow]
\mor{S5}{M3}{}[\atleft, \injectionarrow]
\mor{Si3}{P1}{}[\atleft, \injectionarrow]
\mor{+Si4}{+P2}{}[\atleft, \injectionarrow]
\mor{hSi5}{hP3}{}[\atleft, \injectionarrow]
\mor{Si3}{+Si4}{$\pi_+$}
\mor{+Si4}{hSi5}{}
\enddc
\fn$$
where as usual we also restricted on the space slice by defining the bundle $\hat \Si$.

We stress that the bundle $\hat P$ has {\it by construction} a trivialization with transition functions in the special form $(\vp_+, \vp_+)$. This is in fact explicitly induced by a trivialization on the bundle ${}^+\<\<P$.
On the bundle $\hat P$ we can globally define the subgroup of gauge transformations in the special form $(\phi_+, \phi_+)$. We shall denote this subgroup as $\Aut({}^+\<\<P)\subset \Aut(\hat P)$ since it is an isomorphic image of the group of all gauge transformations on ${}^+\<\<P$. 

Now if $\te^{ab}_\mu$ is a connection on $\hat P$ we can set
$$
\cases{
&\A^i_A= \frac[1/2]\ep^i{}_{jk}\> \te^{jk}_A +\ga\te^{0i}_A\cr
&\K^i_A= \te^{0i}_A\cr
}
\fn$$
Because of the particular form of the transition functions on $\hat\Si$, by going through what we said above we can consider transformation rules of $(\A^i_A, \K^i_A)$ with respect to the transformation rules in $\Aut({}^+\<\<P)\subset \Aut(\hat P)$.
By simply resorting to \ShowLabel{GoodTransformationRulesBI} we easily prove that  $\A^i_A$ transforms as a $\SU(2)$-connection.
Analogously we can prove $\K^i_A$ to be a $\su(2)$-valued $1$-form.

We stress that the bundles $P$ and $\hat P$ are locally isomorphic. 
Consequently, we can consider either the Hilbert-Einstein action, or the selfdual action, or the Holst action on $\hat P$ obtaining field equations locally equivalent to the corresponding field equations on $P$.

Hence there is no evident local  difference between the model written on $P$ and the model written on $\hat P$. However, when the BI connection is defined starting from $\hat P$ the result is a proper $\SU(2)$-connection on ${}^+\<\Si$.
On ${}^+\<\Si$ we hence have two models: the selfdual and the BI models.
Both are models for a generic $\SU(2)$-connection on ${}^+\<\Si$.
The first is obtained out of a spin connection $\om$ on $P$, while the second is obtained out of a spin connection $\te$ on $\hat P$.

If we are in the specific case in which $P$ and $\hat P$ are globally isomorphic, i.e.~$P$ allows a (global) $\SU(2)$-reduction, then $\hat P$ is to be understood as exhibiting such reduction: of course in this case there is a one-to-one correspondence between connections on $P$ and connections on $\hat P$

\NewSection{Spacetime Interpretation of the BI Connection}

Notice that in the previous Section we never resorted to the triviality of the bundle ${}^+\<\Si$.
In fact the construction works perfectly also at spacetime level by defining a $\SU(2)$-connection
directly on ${}^+\<\<P$, i.e.:
$$
\cases{
&\A^i_\mu= \frac[1/2]\ep^i{}_{jk}\> \te^{jk}_\mu +\ga\te^{0i}_\mu\cr
&\K^i_\mu= \te^{0i}_\mu\cr
}
\fn$$
 In some sense $\A^i_\mu$ does here provide a spacetime counterpart to the usual spatial BI connection $\A^i_A$. The derivation for transformation rules (with respect to the $\SU(2)$-gauge transformations) of these fields is very similar to what we have done in the previous Section.
 The BI connection $\A^i_\mu$ is in fact a $\SU(2)$-connection on ${}^+\<\<P$; it is defined  out of a spin connection on $\hat P$ and it restricts to the usual BI connection on ${}^+\<\Si$.

Samuel (see \ref{Samuel}) provided an argument to claim that {\it the Barbero's connection cannot be interpreted as a spacetime connection}.
Of course it is difficult to precisely and rigorously determine what was exactly meant there by {\it spacetime interpretation}, while \ref{gr-qc/0404018} is more explicit in reporting Samuel's paper, claiming that it is impossible to obtain the Barbero-Immirzi connection as a restriction of a suitable spacetime connection.
Despite we agree with Thiemann who refers (see \ref{Thie2006}) to the problem as an {\it aesthetical} one, meaning that
it would not spoil the mathematical consistency of the theory, we believe that a precise understanding of the geometric origin of fields provides better insight on the structure of the theory 
 
We precisely proved above that the BI $\SU(2)$-connection on ${}^+\<\Si$ is obtained by restricting the spacetime $\SU(2)$-connection $\A^i_\mu$ defined on ${}^+\<\<P$; however, $\A^i_A$ is not the restriction of a spacetime {\it spin} connection, of course.
Whether this responds or contradicts Samuel's claim is something we leave to the reader consideration since, in any case, it has no crucial importance here.
What is important, however, is that there is for sure a spacetime interpretation of the appropriate global form of the BI connection.

We think, instead, that  it is instructive to try showing how Samuel's counterexample fits into our framework.
In Samuel's example Minkowski spacetime was considered with two different slicings;
one is the usual $t=c$ slicing, while the other is obtained by a pointwise spin transformation (obtaining some sort of hyperbolic slices).
The two slicings are defined so that there exists a particular loop $\al$ lying on a slice in both slicings.

A frame $\hat e_a$ is choosen to be adapted to the first slicing and, by means of the pointwise spin transformation, a new frame $e_a$ adapted to the second slicing is obtained.
Then the two frames induce two spin connections which in turn define two different BI connections, each adapted to one slicing.
These two spin connections are of course connected by a spin transformation (related precisely to the pointwise spin transformation used for the frames). The argument ends by computing the trace of the holonomy along $\al$ with respect to the two BI connections so obtained. The result does in fact depend on the slicing while the trace of the holonomy of a spacetime connection is expected to be independent of the slicing.
The bundles involved in  the constructions are all trivial and $P$ coincides with its counterpart $\hat P$.
However, the spin transformation used is locally determined by a pointwise element of $\Spin(4)$.
We stress that such transformation is not in the special form $(\vp_+, \vp_+)$.
In our construction the BI connection on ${}^+\<\<P$ is a $\SU(2)$-connection; 
hence the two spin connections presented in \ref{Samuel}  are gauge equivalent with respect to the gauge group $\Spin(4)$, while the corresponding BI connections are not gauge equivalent 
with respect to the smaller gauge group $\SU(2)$.

In other words, the BI connection is a $\SU(2)$-object, not a $\Spin(4)$-object.
One of the best perspectives to look at the BI framework is exactly the need to provide  a $\SU(2)$-formulation of GR already at spacetime level, variously dropping or using the antiselfdual part of the spin group.
Again whether this provides a satisfactory spacetime interpretation of the BI connection is left to the reader.
In any case, we believe that it explains why one should not expect the same trace-holonomy for the two spacetime connections; the two connections are not gauge equivalent with respect to their gauge group $\SU(2)$.

\NewSection{Lorentzian case}

Most of the results obtained for an Euclidean connection do in fact rely on the special form 
\ShowLabel{SU2SpecialForm}.

In the Lorentzian case the relevant spin group is $\Spin(3,1)\simeq \SL(2, \C)$ which is in fact the ``complexification'' of the group $\SU(2)$. Thus we have a canonical group embedding
$\iota:\SU(2)\arr \Spin(3,1)$ exhibiting $\SU(2)$ as a real section of $\Spin(3,1)$.

We do in fact re-obtain results similar  to the Euclidean case considered above by noticing that, under very reasonable assumptions, the spin bundle $P$ does in fact allow a $\SU(2)$-reduction relative to the group homomorphism $\iota:\SU(2)\arr \Spin(3,1)$:
$$
\begindc{\commdiag}[1]
\obj(110,80)[+P2]{$^+P$}
\obj(180,80)[hP3]{$P$}
\obj(70,50)[+Si4]{$^+\Si$}
\obj(140,50)[hSi5]{$\Si$}
\obj(110,30)[M2]{$M$}
\obj(180,30)[M3]{$M$}
\obj(70,0)[S4]{$S$}
\obj(140,0)[S5]{$S$}
\mor{+P2}{M2}{}
\mor{hP3}{M3}{}
\mor{+Si4}{S4}{}
\mor{hSi5}{S5}{}
\mor{+P2}{hP3}{}
\mor{M2}{M3}{}[\atleft, \solidline] \mor(110,33)(180,33){}[\atleft, \solidline]
\mor{S4}{S5}{}[\atleft, \solidline] \mor(70,3)(140,3){}[\atleft, \solidline]
\mor{S4}{M2}{}[\atleft, \injectionarrow]
\mor{S5}{M3}{}[\atleft, \injectionarrow]
\mor{+Si4}{+P2}{}[\atleft, \injectionarrow]
\mor{hSi5}{hP3}{}[\atleft, \injectionarrow]
\mor{+Si4}{hSi5}{}
\enddc
\fn$$

In dimension four the reduction is related to the third Stiefel-Whitney class of $M$ (see \ref{Antonsen} and references quoted therein).
Such class is trivial when both the first and the second Stiefel-Whitney classes are trivial (which can be proved by using Steenrod square operators in cohomology; see \ref{Milnor}).
On  the other hand, the first and second Stiefel-Whitney classes of $M$ are already assumed to be trivial to allow spin structures on $M$ (which are needed to define spinors, which of course exist in our spacetime).

As in the Euclidean case gauge transformations on ${}^+\<\<P$ induce a subgroup $\Aut({}^+\<\<P)\subset \Aut(P)$ of gauge transformations on $P$. Then the elements $\si$ in  this subgroup induce Lorentz transformations which are in the simplified form
$$
\ell(\si)=\(\matrix{
1 & 0 \cr
0 & \la(\si)\cr
}\)
\fn$$
This allows to prove easily that 
$$
\cases{
&\A^i_\mu= \frac[1/2]\ep^i{}_{jk}\> \te^{jk}_\mu +\ga\te^{0i}_\mu\cr
&\K^i_\mu= \te^{0i}_\mu\cr
}
\fn$$
are again well-behaving fields, regardless of the signature and without complexifications of the connection $\te$.

Some differences between the Euclidean and the Lorentzian formulation still exist; 
for example we loose the beatifully explicit construction of the $\SU(2)$-reduction in the Euclidean case (which is here just proven to exist). 
Then the slice $S\subset M$ has to be spacelike with respect to the frame which is part of the field configuration (if not the residual triad field defined on $S$ would not be a $\SU(2)$-field itself; see \ref{OurAshtekar}).
This means that one fixes a slice and then restricts the configurations allowing just frames for which the slice is spacelike. Different configurations are obtained for different choices of the initial slice.
A single slice covers a whole set of possible configurations so that, with a countable number of choices,
all possible configurations are obtained.
However, each local framework defined in this way is real and geometrically well-defined.

\

\NewSection{Conclusions and Perspectives}

We have provided a global geometric framework to introduce the BI connection and understand its global properties.
We have also shown that the BI connection does in fact appear as the restriction of a global $\SU(2)$-connection defined on the whole spacetime.
The construction does not rely on the possible triviality of the principal bundle which encodes the gauge structure  of the model nor it resorts to gauge fixings which would spoil manifest gauge covariance.
On the contrary, the construction relies on the existence of a $\SU(2)$-reduction which is the correct mathematical structure to be considered.

We believe that this framework might help to investigate the global gauge structure of the theory and the relations among different gauge groups $	\Spin(4)$, $\Spin(1,3)$, $\SU(2)$ which appear in LQG.
These groups encode the covariance properties of GR and a better control on their mutal relations might provide a suitable framework to clarify the covariance issues which are sometimes still under 
discussion in LQG.

Finally, the spacetime interpretations of the objects appearing in LQG might help in clarifying the issues connected to the semiclassical limits of LQG. 

Future investigations will be devoted to clarify the role of the field $\K^i_\mu$.
In fact, thanks to the spacetime fields here introduced $(\A^i_\mu, \K^i_\mu)$, one can
pull-back the Holst's action to ${}^+\<\<P$ obtaining a good $\SU(2)$ formulation for GR at a spacetime level. The results are in progress and they will form the core of a forthcoming paper on this subject.

\NewAppendix{\SelfDualEquivalenceAPP}{Transformation Rules of Different Fields in GR}

We shall here list the transformation rules of the objects which have been used to provide different descriptions of the GR.

The spin connection on $P$ is described by coefficients $\om^{ab}_\mu$ which transform as
shown in \ShowLabel{SpinConnectionTranformationRules}.

The bundle ${}^+\<\<P$ is a principal bundle with group $\SU(2)\equiv \Spin(3)$; hence a connection on it is described by coefficients $\om^{ij}_\mu(x)$ ($i,j=1,2,3$ are skewsymmetric indices) transforming as
$$
\om'{}^{ij}_\mu=\bar J^\nu_\mu \la^i_m\>\(\om^{mn}_\nu\> \la^j_n+ \di_\nu \bar\la^m_n\eta^{nj}\)
\fn$$
All along the paper, connections on ${}^+\<\<P$ have been represented by coefficients
$\om^{k}_\mu=\frac[1/2]\ep^k{}_{ij} \om^{ij}_\mu$. They transform as
$$
\eqalign{
\om'{}^{k}_\mu=& \frac[1/2]\ep^k{}_{ij} \om'{}^{ij}_\mu= 
\frac[1/2]\ep^k{}_{ij}\bar J^\nu_\mu \la^i_m\>\(\om^{mn}_\nu\> \la^j_n+ \di_\nu \bar\la^m_n\eta^{nj}\)=\cr
=&
\bar J^\nu_\mu \(\la^i_j\>\om^{j}_\nu-\frac[1/2]\ep^{kj}{}_{i} \la^i_l\di_\nu \bar\la^l_j\)
}
\fl{A2}$$
The information contained in the spin connection $\om^{ab}_\mu$ can be expressed in a number of equivalent ways.

\NewSubSection{Barbero-Immirzi connection}

The spin connection is described by $24=6\times 4$ functions.
We can split the same information as $(A^i_\mu, K^i_\mu)$ which are $(3\times 4) \oplus (3\times 4)$ functions; see \ShowLabel{BIConnection} and \ShowLabel{BIK}. The map $\om^{ab}_\mu\mapsto (A^i_\mu, K^i_\mu)$ is invertible for every $\ga\not=0$ and the new objects have a global meaning if the spin bundle $P$ has a $\SU(2)$-reduction; see \ShowLabel{GoodTransformationRulesBI} for transformation rules.

\NewSubSection{Selfdual Connection}

The (anti)selfdual connection ${}^\pm\<\om^i_\mu$ are again $(3\times 4) \oplus (3\times 4)$ functions.
The transformations rules of ${}^\pm\<\om^i_\mu$ are obtained in general (without resorting to a $\SU(2)$-reduction of $P$). 

Let us consider the transformations rules of  $\om^{ab}_\mu$ with respect to a gauge transformation $(S_+, S_-)$ and let us  denote by $\la$ the $\SO(3)$ transformation induced by $S_\pm$;
then the transformation rules for ${}^\pm\<\om^{i}_\mu$ are (see \ShowLabel{A2})
$$
{}^\pm\<\om'{}^{i}_\mu=\bar J^\nu_\mu \(\la^i_j\>{}^\pm\<\om^{j}_\nu-\frac[1/2]\ep^{kj}{}_{i} \la^i_l\\di_\nu \bar\la^l_j\)
\fn$$
The check of the necessary identities was performed by using MapleTensor package.

\NewSubSection{Deformed connection}

In the analysis of Holst's action (which is not carried out here) is useful to introduce
$$
{}^\ga\<\om^{ab}_\mu = \frac[1/2] \ep\^a\^b{}_{cd} \om^{cd}_\mu +\ga\om^{ab}_\mu 
\fn$$ 
These are again $6\times 4$ independent functions. They are not in general the components of a $\Spin(4)$-connection.
In the Euclidean case, the transformations are invertible for  $\ga\not=\pm 1$ (while in the Lorentzian case they are invertible for any real nonzero value; of course invertibility would be lost for $\ga=\pm i$).
Of course for $\ga=\pm1$ the Euclidean (anti)selfdual connection is obtained.

We also have
$$
\cases{
&{}^\ga\<\om^{0i}_\mu = A^i_\mu\cr
&{}^\ga\<\om^{ij}_\mu = \ep^{ij}{}_{k} \(\ga A^k_\mu +(1-\ga^2) K^k_\mu\)\cr
}
\fn$$ 
Notice how the selfdual formalism is dealt differently with respect to the BI case. 
In the (anti)selfdual case one uses the equivalent variables ${}^\pm\<\om^i_\mu$, while
in the Holst's case one uses ${}^\ga\<\om^{ab}_\mu $ (for a single value of $\ga$) which are in fact non-degenerate for $\ga\not=\pm 1$.
Finally, in the BI setting we use $(A^i_\mu, K^i_\mu)$  as variables.
The ultimate reason for different strategies is due to algebraic degeneracy of the (anti)selfdual case.

\Acknowledgements

One of us (LF) thanks Google for pointing out so efficiently the paper \ref{Antonsen}.

This work is partially supported by MIUR: PRIN 2005 on ``Leggi di conservazione e termodinamica in meccanica dei continui e teorie di campo''.  We also acknowledge the contribution of INFN (Iniziativa Specifica NA12) and the local research founds of Dipartimento di Matematica of Torino University.

\ShowBiblio

\end